\documentclass[aps,prl,twocolumn,superscriptaddress,showpacs,floatfix,footinbib]{revtex4-2}
\usepackage{graphicx}
\usepackage{dcolumn}
\usepackage{bm}
\usepackage[hidelinks]{hyperref}
\usepackage{graphicx,amsmath, amssymb} 
\usepackage{amsmath}
\usepackage{braket}
\usepackage{xcolor}
\usepackage{pifont}
\usepackage{layouts}   
\usepackage{tikz}
\newcommand\abs[1]{\left|#1\right|}
\renewcommand{\figureautorefname}{Fig.}
\usepackage[]{bibunits}
\begin{document}
\makeatletter
\newcommand{\silentcite}[1]{%
    \@bsphack
    \protected@write\@auxout{}%
        {\string\citation{#1}}%
    \@esphack%
}
\title{Spiral renormalization group flow and universal entanglement spectrum of the non-Hermitian 5-state Potts model}
\author{Vic Vander Linden}
\affiliation{Institute for Theoretical Physics Amsterdam, University of Amsterdam, P.O. Box 94485, 1090 GL Amsterdam, The Netherlands}
\affiliation{Department of Physics and Astronomy, Ghent University, Krijgslaan 281, 9000 Gent, Belgium}
\author{Boris De Vos}
\affiliation{Department of Physics and Astronomy, Ghent University, Krijgslaan 281, 9000 Gent, Belgium}
\author{Kevin Vervoort}
\affiliation{Department of Physics and Astronomy, Ghent University, Krijgslaan 281, 9000 Gent, Belgium}
\author{Frank Verstraete}
\affiliation{Department of Physics and Astronomy, Ghent University, Krijgslaan 281, 9000 Gent, Belgium}
\affiliation{Department of Applied Mathematics and Theoretical Physics, University of Cambridge,\\ Wilberforce Road, Cambridge, CB3 0WA, United Kingdom}
\author{Atsushi Ueda}
\affiliation{Department of Physics and Astronomy, Ghent University, Krijgslaan 281, 9000 Gent, Belgium}

\date{\today}
\begin{abstract}
The quantum $5$-state Potts model is known to possess a perturbative description using complex conformal field theory (CCFT), the analytic continuation of ``theory space" to a complex plane. To study the corresponding complex fixed point on the lattice, the model must be deformed by an additional non-Hermitian term due to its complex coefficient $\lambda$. Although the variational principle breaks down in this case, we demonstrate that tensor network algorithms are still capable of simulating these non-Hermitian theories. We access system sizes up to $L = 28$, which enable the observation of the theoretically predicted spiral flow of the running couplings. Moreover, we reconstruct the full boundary CCFT spectrum through the entanglement Hamiltonian encoded in the ground state. Our work demonstrates how tensor networks are the correct approach to capturing the approximate conformal invariance of weakly first-order phase transitions.
  
\end{abstract}

\maketitle
\begin{figure}[t!]
    \centering
      {\includegraphics[trim={0 0.cm 0 0},clip,width=7.cm]{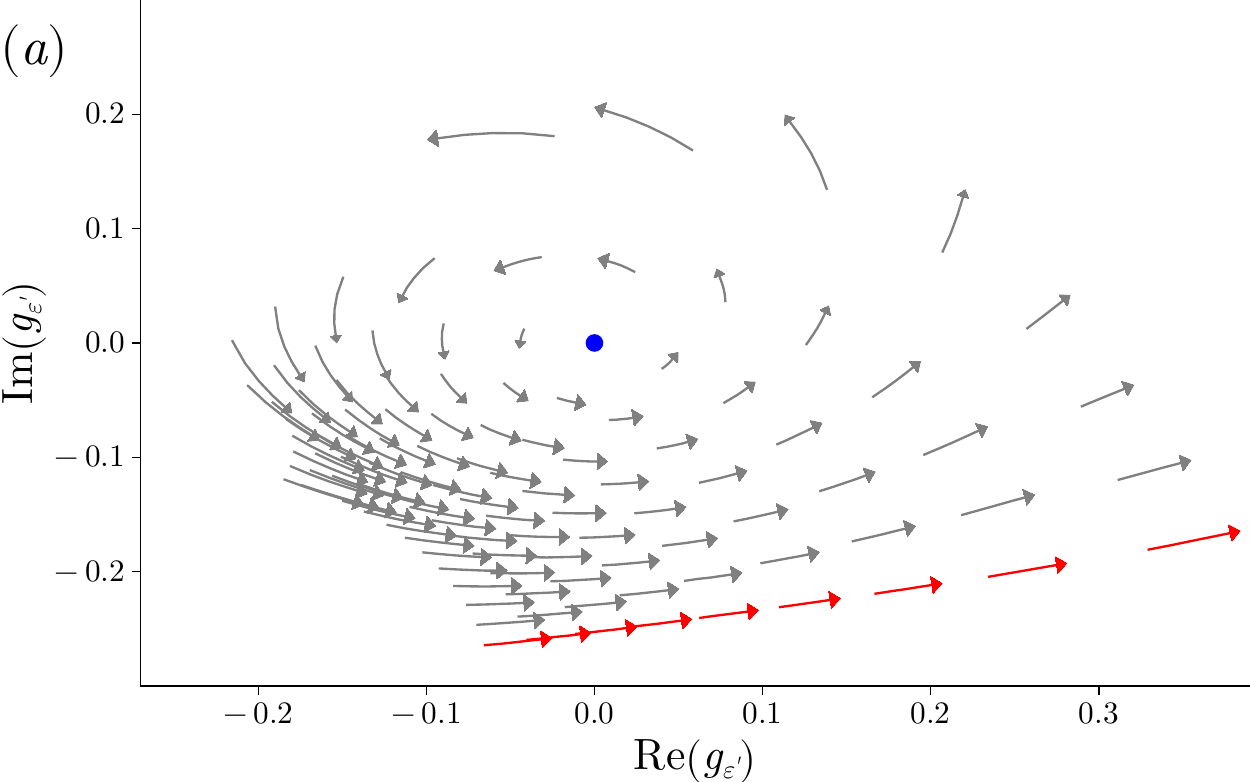}}\par
    {\includegraphics[trim={0 0.cm 0 0},clip,width=7.2cm]{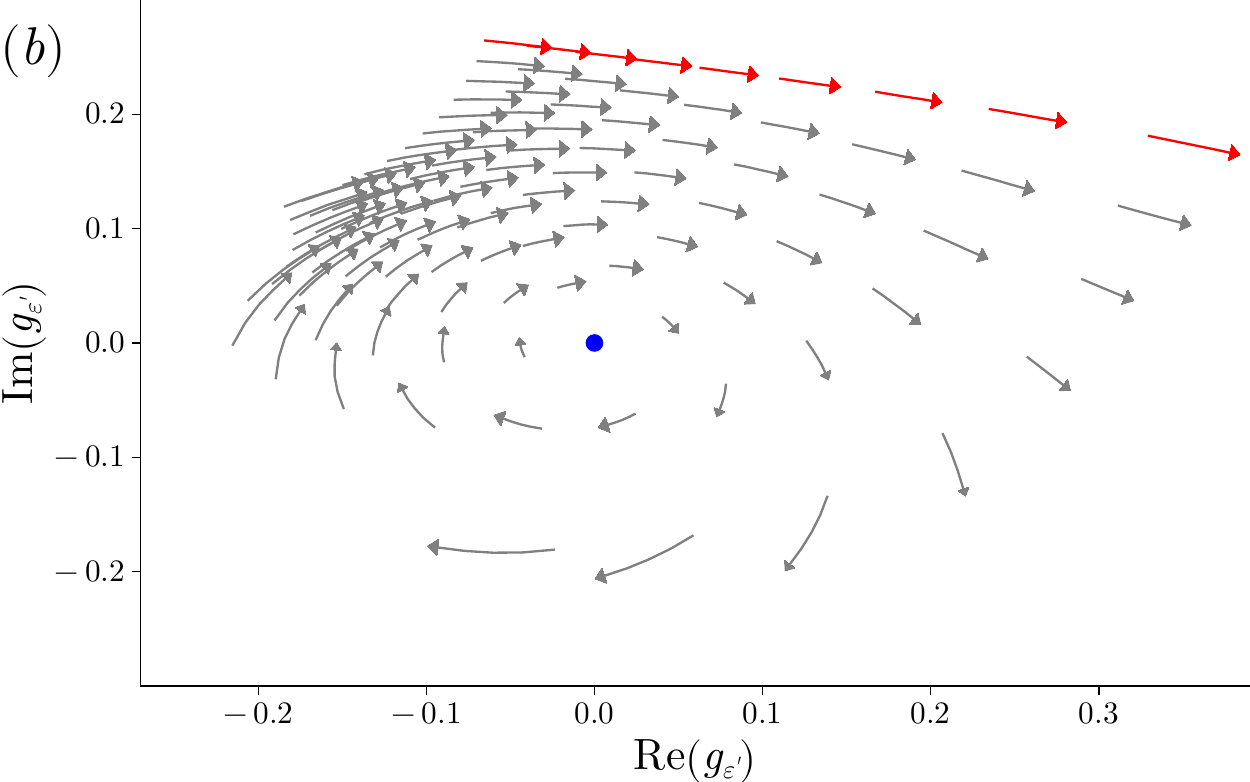}} \par
    \caption{The flow of the running coupling parameter $g_{\varepsilon'}$ acquired through the cost function fitting \cite{Tang2024,supplementary} (a) for $\mathcal{\overline{C}}$ and (b) for $\mathcal{C}$ around their respective fixed points for system sizes $L=16-24$. The perturbation parameter at $\lambda_c$ is denoted by a blue dot, while red arrows show the walking behavior of values on the real axis.}
    \label{fig:rgflowplot}
\end{figure}
\begin{figure}[t!]
    {\includegraphics[trim={0cm 0.4cm 0cm 1.2cm},clip,width=0.48\textwidth]{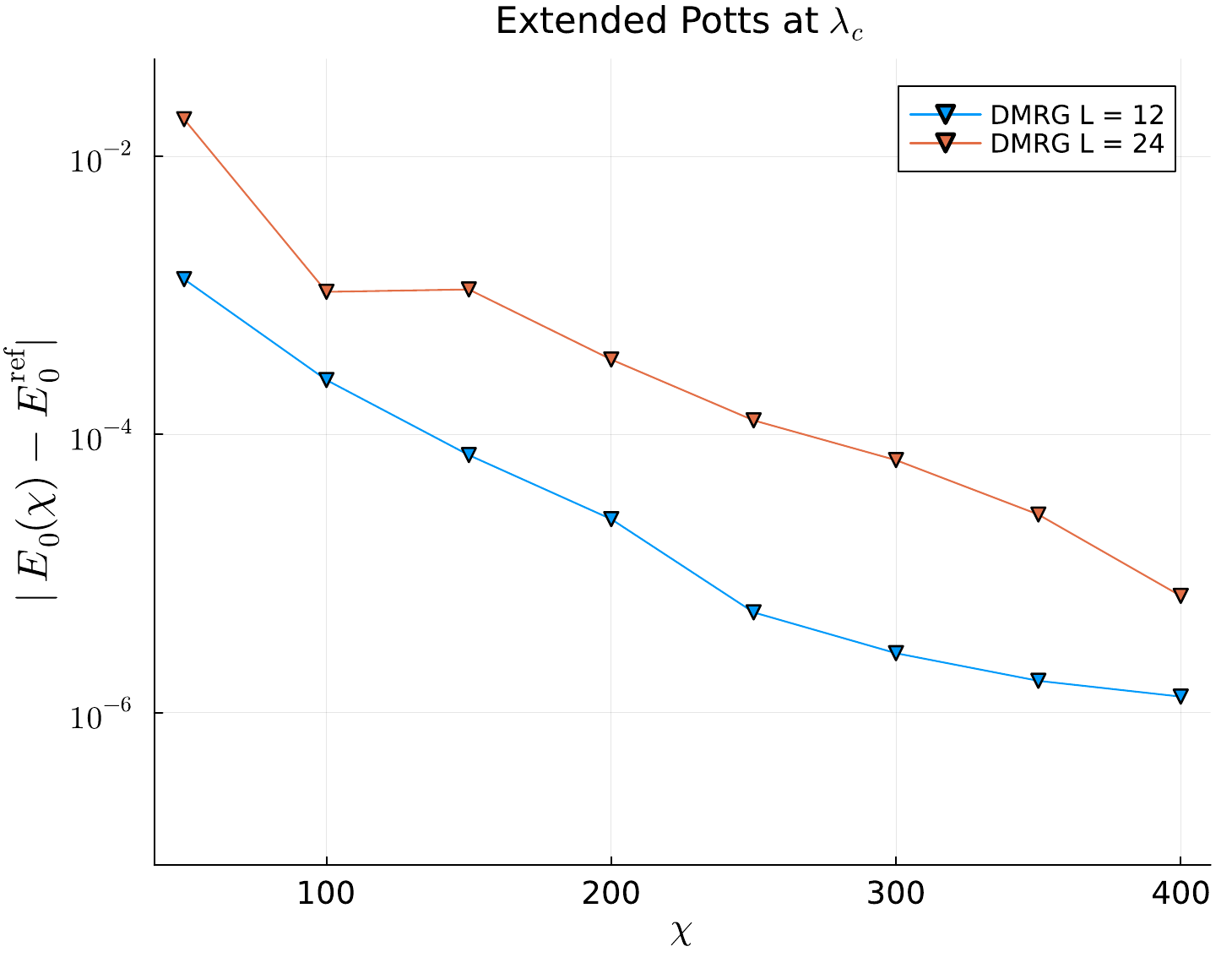}\label{scaling1}}
     \makeatletter\long\def\@ifdim#1#2#3{#2}\makeatother
    \caption{The difference between the ground state energy $E_0$ and a reference energy $E^{\text{ref}}_0$ at the critical point $\lambda_c$  with increasing bond dimension. For $ L = 12$ the reference is the ED ground state and for $ L = 24$ it is $E_0( \chi = 450)$.}
    \label{fig: dif}
\end{figure}
\begin{figure}[tb]
    \centering
    \includegraphics[trim={0 0.7cm 0 0},clip,width=0.2222\textwidth]{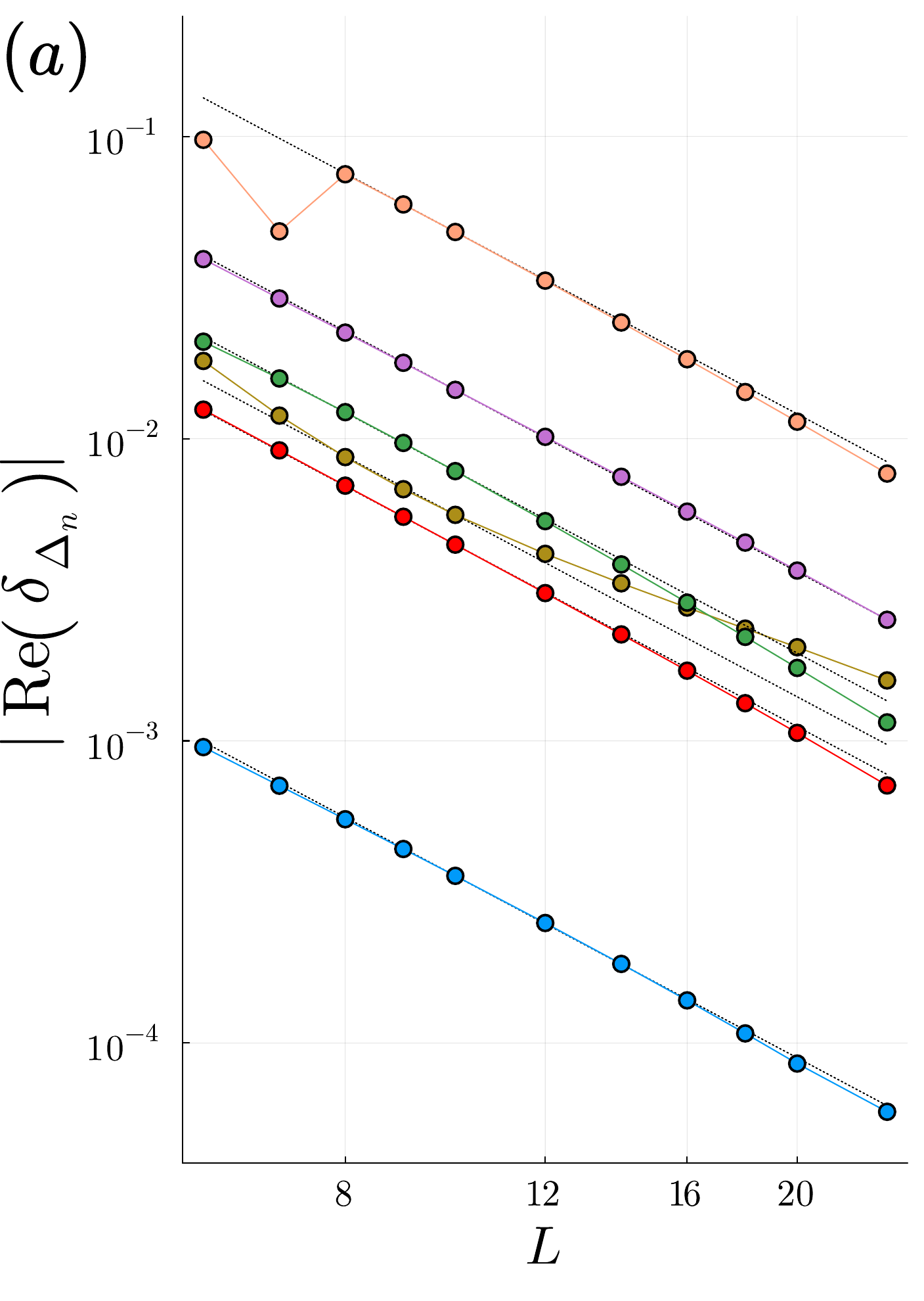}
   \includegraphics[trim={0 0.7cm 0 0},clip,width=0.2548\textwidth]{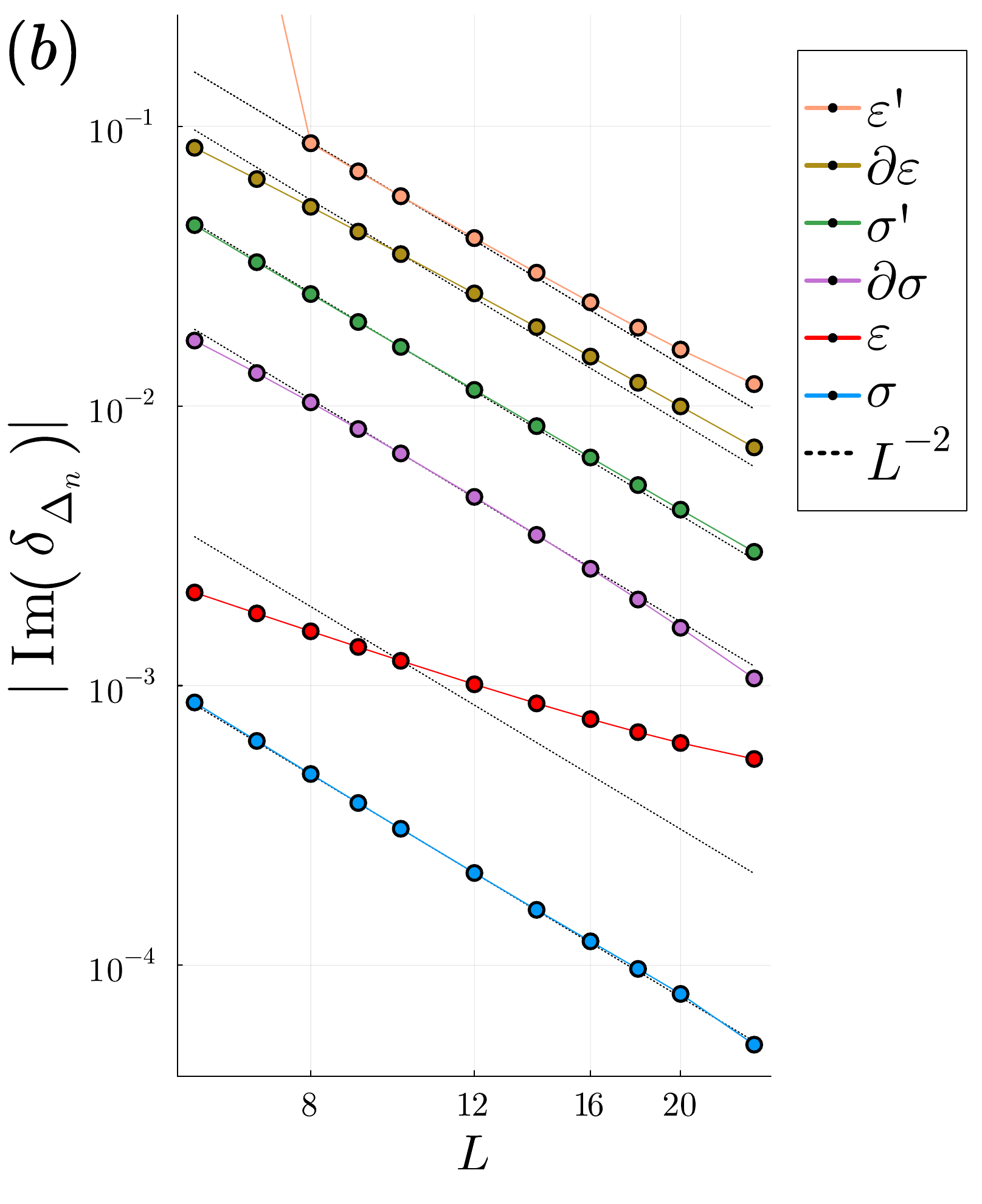}
    \caption{The $1/L^2$ scaling of the difference between the theoretical and estimated scaling dimensions Eq. \eqref{eq:Deltadeviation} at $\lambda_{c}  = 0.0788 + 0.0603i$ for (a) the real part and (b) the imaginary part.}
    \label{fig:lambdacscaling}
\end{figure}
\begin{figure*}[tb]
    \centering
     \includegraphics[trim={9cm 2.5cm 9cm 0cm},clip,width=0.495\textwidth]{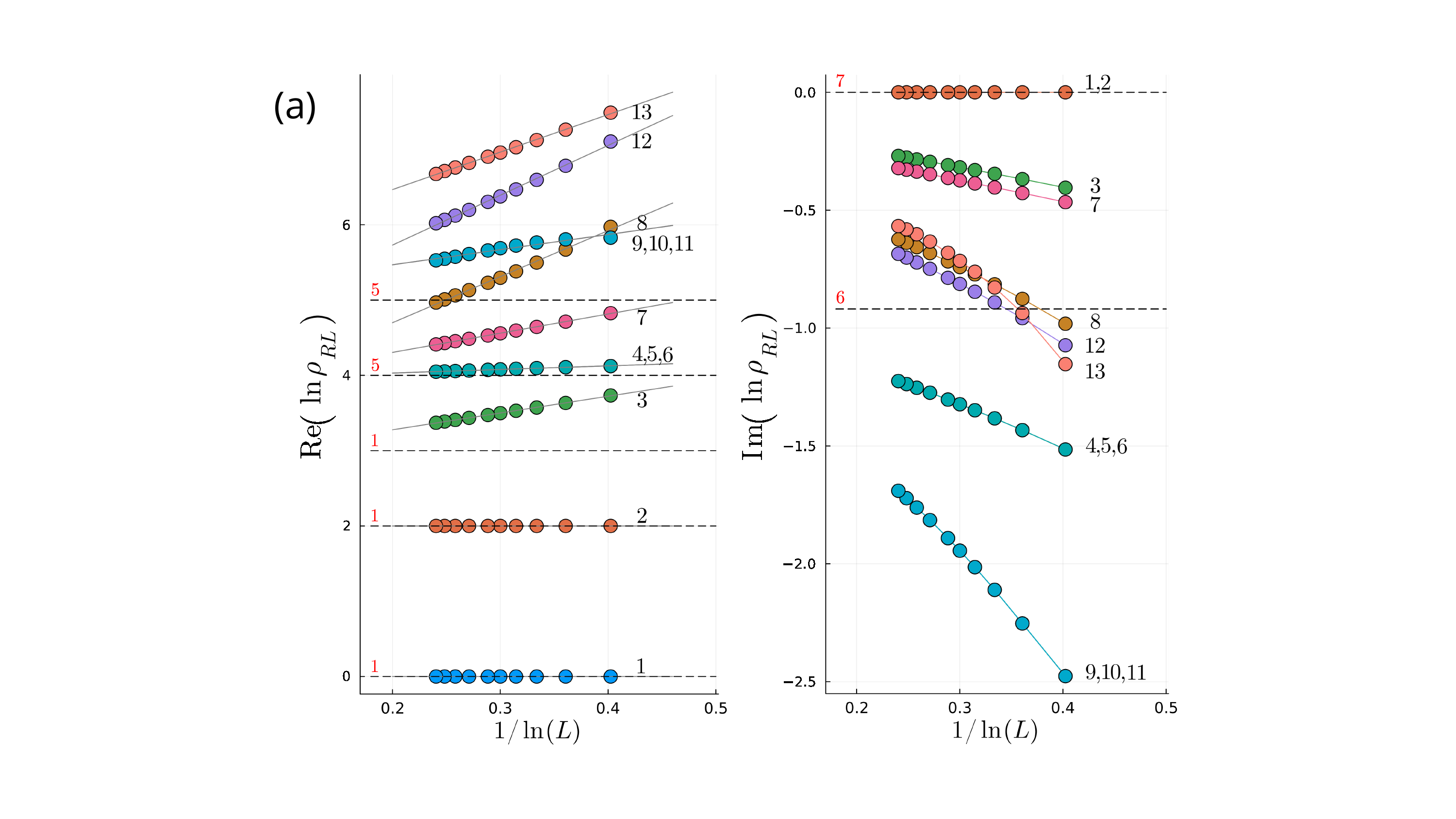}
    \includegraphics[trim={9cm 2.5cm 9cm 0cm},clip,width=0.495\textwidth]{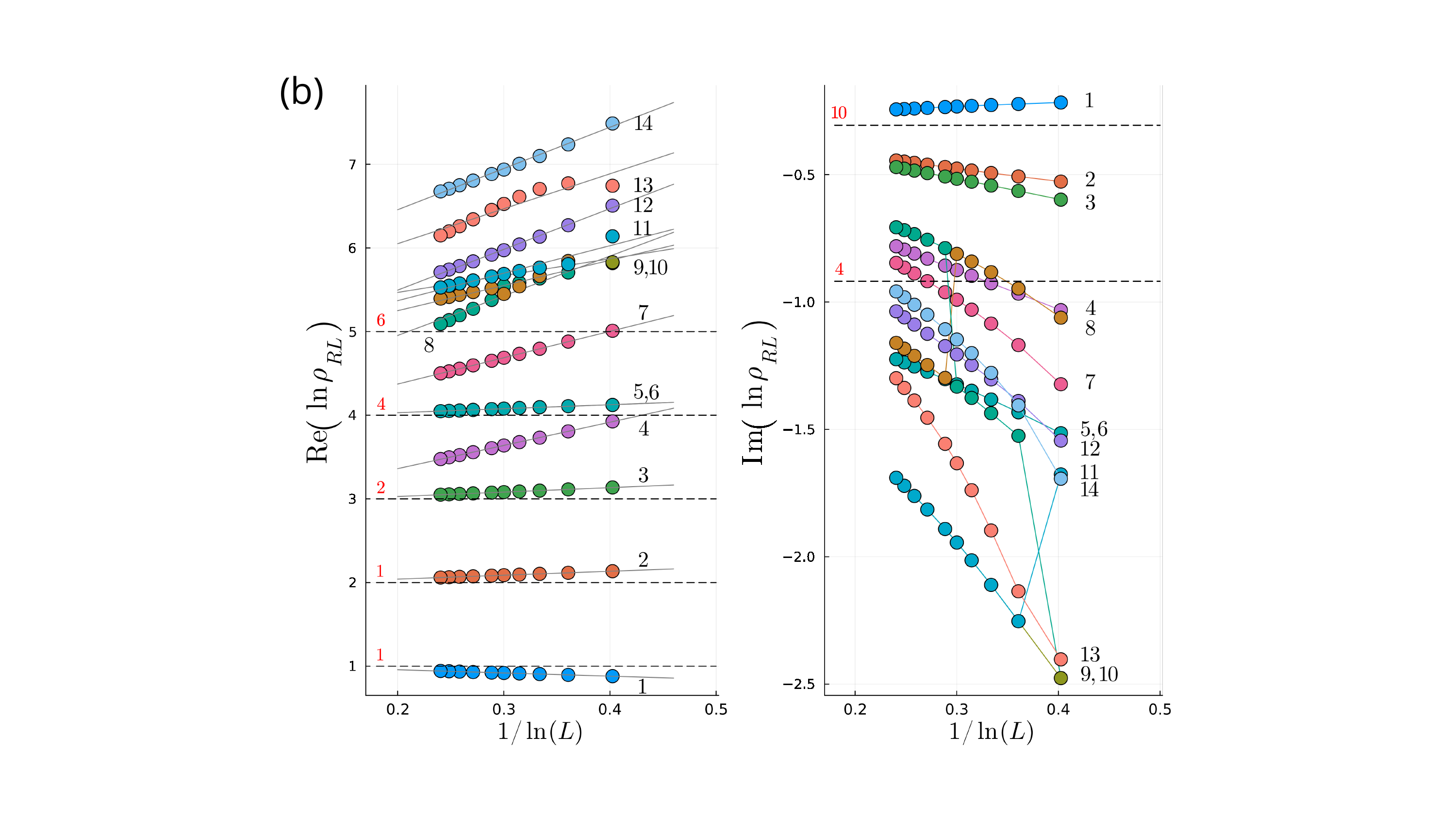}
    \caption{The entanglement spectrum of the ground state in the (a) zero sector and (b) a charged sector of $\mathbb{Z}_5$ for the deformed 5-state Potts Hamiltonian at $\lambda_c = 0.0788 + 0.0603i$ for system sizes $L = 12 - 64$ and periodic boundary conditions. The dashed lines and red numbers are the theoretical energy levels and their degeneracies, respectively \cite{Tang:2025bju}. We fit the data to their theoretical values for the first two energy levels of the zero sector.}
    \label{fig:BCFT_entanglementspectrum}
\end{figure*}

\begin{bibunit}[apsrev4-2]
\indent \emph{Introduction—}The renormalization group (RG) is a fundamental framework for understanding and classifying continuous phase transitions. At fixed points of the RG flow, the theory becomes scale invariant and is often described by a conformal field theory (CFT). Intriguingly, some first-order phase transitions --- particularly \emph{weakly first-order} ones --- exhibit approximate scale invariance over an intermediate length scale. In this regime, the RG flow slows dramatically, a phenomenon known as \emph{walking}, which has been discussed in various contexts~\cite{Kaplan2009,Corboz2013, Xi2023, Koga2000, Lee2019,Naichuk_2024,Ma2019,kumar2025}. This walking behavior suggests the presence of nearby fixed points that influence the flow without being directly accessible. This leads to a compelling scenario: the slow RG flow can be viewed as a crossover governed by a pair of \emph{complex fixed points}, which are analytic continuations of real fixed points into the complex coupling space. These fixed points are not in the theory space of the original real walking parameter and hence imply a first-order phase transition. Nevertheless, the shadow of these invisible fixed points imprints quasi-critical behavior on the system~\cite{Ma2019, Gorbenko2}.
\\
\indent The $1+1$D quantum $Q$-state Potts model is a paradigmatic example of this phenomenon. This model is known to undergo a continuous phase transition for $ Q \leq 4$ and a first-order transition for $Q > 4$, consistent with the two real fixed points for $Q \leq 4$, a critical and tricritical one, which overlap at $Q = 4$. The weakly first-order nature of the $Q = 5$ case has been conjectured to originate from a \textit{complex conformal field theory} (CCFT) at the complex fixed points that governs the walking regime~\cite{Gorbenko1, Gorbenko2}. This scenario has been theoretically proposed based on the analytic continuation of Coulomb gas techniques. Recently, a lattice regularization of the 5-state Potts CCFT was found by extending the Potts Hamiltonian with a non-Hermitian term, verifying the existence of the complex fixed points $\mathcal{C}$ and $\overline{\mathcal{C}}$ \cite{Tang2024}. The deformed 5-state Potts model is given by
\begin{align} \label{eq:totalham}
    \hat{H} = &\sum_i \sum_{k=1}^{Q-1} \left[J(\hat{Z}^\dag_i \hat{Z}_{i+1})^k + h \hat{X}_i^k \right] + \lambda \hat{H}_1,\\
   \hat{H}_1 = &\sum_i \sum_{k_1=1,k_2=1}^{Q-1} [ (\hat{X}_i^{k_1} + \hat{X}_{i+1}^{k_1})(\hat{Z}_i^{\dag} \hat{Z}_{i+1})^{k_2} \label{eq:extensionterm}\\
    \nonumber & + (\hat{Z}_i^{\dag} \hat{Z}_{i+1})^{k_1} (\hat{X}_i^{k_2} + \hat{X}_{i+1}^{k_2}) ],
\end{align} 
where $\hat{X}$ is the spin shift operator which acts on the $Q$ states labeled $\ket{n}, n = 0, \hdots, Q-1$ as $\hat{X} \ket{n} = \ket{(n+1)\mod Q}$ and $\hat{Z}$ the phase operator which acts as $\hat{Z} \ket{n} = e^{\frac{2\pi n i}{Q}} \ket{n}$. The critical value $\lambda_c \in \mathbb{C}$ was found by exact diagonalization methods to simulate the complex fixed points \cite{Tang2024}. The RG flow around these fixed points, as predicted in~\cite{Gorbenko2} and verified in Fig. \ref{fig:rgflowplot}, materializes as a logarithmic spiral in the complex theory space, along with a straight line representing the walking behavior of the original 5-state Potts model. \\
\indent In this Letter, we further explore this CCFT scenario directly from the lattice using the tensor network framework \cite{Cirac2021, schollwock2005density,verstraete2008matrix} which allow for bigger system sizes than allowed by exact diagonalization. This has enabled us to uncover the spiral trajectory of the running couplings. A priori, a non-Hermitian adaptation of tensor network algorithms is required to properly capture the left and right eigenvectors of non-Hermitian Hamiltonians \cite{Ashida_2020}. However, we demonstrate how the walking behavior of the Hermitian theory renders the non-Hermitian deformation sufficiently small, allowing unadapted algorithms to capture the complex fixed point accurately. We illustrate this accuracy by predicting a more precise critical value $\lambda_c$ based on finite-size scaling, achieving a more accurate estimate for the bulk conformal data. With this, we reveal the spiral RG flow of the running coupling. In addition, we demonstrate how DMRG still allows us to capture the boundary CCFT through the entanglement Hamiltonian. 
\\
\indent \emph{Tensor networks}---We define the ground state wavefunction of the deformed Potts Hamiltonian as being the eigenstate corresponding to the eigenvalue with the smallest real part, and assume that the Schmidt coefficients decay fast enough such that it can be represented as a matrix product state (MPS) \cite{Verstraete_2006}. In particular, we impose the $\mathbb{Z}_5$ subgroup symmetry of the Hamiltonian for computational efficiency and to identify the charges of the CCFT primary operators easily. We use an MPS ansatz with open BC (and a Hamiltonian with periodic BC, hence with one long-ranged interaction term) such that we can employ the standard density matrix renormalization group (DMRG) \cite{White1992}. The excited states are determined by an algorithm that targets the quasiparticle excitations \cite{Vandamme2021, Haegeman2013}. We find that the ground state and excitation energies deviate slightly from the values obtained by exact diagonalization (ED) as shown in Fig.~\ref{fig: dif}. However, the errors decrease with increasing bond dimension with an algorithmic tolerance $10^{-6}$. This can be understood from the fact that the Hamiltonian~\eqref{eq:totalham} contains only a small portion of non-Hermiticity, thereby rendering the MPS truncation almost optimal. In the following, we use $\chi = 400$ and $\chi = 600$ for the ground state and excitation calculations. We also confirmed that the errors are in the same order as the algorithmic tolerance in the larger systems.
Consequently, we can achieve accurate results up to system sizes $L = 24$, while ensuring that what remains are the finite-size effects and that second-order terms stay negligible \cite{Pirvu2012}. In particular, we use larger system sizes beyond the reach of ED methods up to $L=64$ for computing the entanglement spectrum. Employing larger systems is crucial, for instance, in Fig.~\ref{fig:BCFT_entanglementspectrum}, where the maximal reach of ED methods $L\sim 12$ barely captures the right physics due to the logarithmic finite-size effects. The relatively good performance of the MPS algorithms on non-Hermitian models in capturing the entanglement can also be explained by the small non-Hermiticity. It retains the same ordering as the true entanglement structure. Consequently, as further analyzed in \cite{supplementary}, the algorithms only experience slight systematic errors in the imaginary results, a known problem \cite{Shimizu2025}. We use the Julia packages TensorKit.jl and MPSKit.jl to perform the numerical simulations \cite{tensorkit, mpskit}. \\ \\
\indent \emph{Conformal data}---To increase the precision of the fixed point $\lambda_c$, we consider leading irrelevant perturbations from the operators $\varepsilon'$, $T \overline{T}$, $T^2$ and $\overline{T}^2$ allowed by the symmetries of the Hamiltonian  The (spatial) symmetry requirements leave only $\partial^4 \mathbb{I},\overline{\partial}^4 \mathbb{I}$ and $\Box \varepsilon'$ which have vanishing one-point functions or orthogonal Verma modules to our group of operators \cite{Cardy1986}.
We simulate the Hamiltonian with PBC to exploit radial quantization. Consequently, at the critical point where $g_{\varepsilon'} = 0$, the ground state energy and energy gaps depend on the central charge $c$ and scaling dimensions $\Delta_n$ of the CCFT according to \cite{affleck1986,blote1986}
\begin{align}\label{eq:E0Ensubleading}
    \begin{aligned}
         \frac{E_0}{L} &= f - \frac{\pi v c }{6} x^2 + a x^4, \\ 
         (E_{n} - E_0)L &= 2\pi v \Delta_n + \tilde{a}_n x^2,
    \end{aligned}
\end{align}
 where $x = 1/L$, $v$ is the speed of light, $f$ is the ground state energy density and $a,\tilde{a}_n$ take into account the leading irrelevant perturbations. For each point $\lambda$ in theory space, we first recover the non-universal parameter $v$ by plugging in the theoretical central charge $c$ and ground state energies for $L = 8-14$. We extrapolate the fit of Eq. \eqref{eq:E0Ensubleading} for all pairs of subsequent data points to $L \to + \infty$ to obtain $v \approx 2.880 -0.705i$. These intermediate system sizes allow us to exclude sub-subleading finite-size corrections, while also avoiding deviations due to the finite bond dimension.  Now we evaluate the deviation between the leading order contribution to $\Delta_n$ in Eq. \eqref{eq:E0Ensubleading} and the theoretical value $\Delta_n^\text{th}$ \cite{Gorbenko2}:
\begin{equation}\label{eq:Deltadeviation}
    \delta_{\Delta_n} = \frac{L}{2 \pi v}(E_{n}-E_{0}) - \Delta_n^\text{th}.
\end{equation} 
By analyzing the scaling of these values for different system sizes, we recover the fixed point as the value for $\lambda$ where we observe the correct $1/L^2$ scaling.\\
\indent The best subleading scaling for the real and imaginary part of the six most relevant operators is visible in \autoref{fig:lambdacscaling}, yielding the critical point $\lambda_c = 0.0788 + 0.0603i$. We find only the imaginary part of the $\varepsilon$ operator to deviate from this expected scaling. This might be attributed to the rather large imaginary part of the OPE coefficient $C_{\varepsilon\varepsilon\varepsilon'}$ \cite{Gorbenko2}. Moreover, for very small deviations, we see that the subleading scaling is highly sensitive to $v$. In particular, we refine the speed of light to $v = 2.8812 + 0.7050i$ to produce a better scaling. Notably, this new fixed point $\lambda_c$ shows a higher precision compared to the previous estimate $\lambda_c^{\text{prev}} = 0.079 + 0.060i$ \cite{Tang2024, supplementary}.
 \\
\indent We now extract the conformal data with the refined critical parameter. To do so, we fit Eq. \eqref{eq:E0Ensubleading} to the fit parameters $\Delta_n$, $f$, $c$, $a$ and $\tilde{a}_n$ for $L = 8-14$. The conformal scaling dimension estimates are visible in \autoref{tab:conformaldata_comparison}. We compare these values with the theoretical predictions found in \cite{Gorbenko2}. Additionally, we include the scaling dimensions extracted through ED as done in \cite{Tang2024} for $\lambda_{c}^\text{prev} = 0.079 + 0.060i$ and $L = 8-13$. We observe that the improved accuracy of our new fixed point yields more accurate estimates for the conformal dimensions. The ED result at $\lambda_{c}^\text{prev}$ only has better accuracy compared to the MPS algorithms at $\lambda_{c}$ for the imaginary parts of the conformal data. This is most likely due to the non-Hermitian MPS ansatz having more untrustworthy low entanglement contributions \cite{supplementary}. Moreover, the tensor network framework deteriorates with increasing scaling dimension due to deviations from the area law. Lastly, the imaginary part of $\Delta_{\varepsilon}$ in \autoref{tab:conformaldata_comparison} was fitted for $1/L^2$, which according to \autoref{fig:lambdacscaling} does not appear to be correct. The scaling which is closest to the theoretical value is $1/L^{1.05}$ with a value of $\mathrm{Im}(\Delta_{\varepsilon}) = - 0.2245(1)i$. It remains unclear whether this observation has a deeper meaning or if it is due to the same simulation artifacts discussed above.
\begin{table}[tb]
\begin{ruledtabular}
    \begin{tabular}{lccc}
      $\,$ & $\text{Complex CFT}$ & $\text{DMRG/QPA at } \lambda_{c}$ &$\text{ED at }$ $\lambda_{c}^\text{prev}$ \\
    \hline
    $c$ & $1.1375 - 0.0211i$ & $1.1376(1) - 0.0212(1)i$    &  $1.1405 - 0.0224i$\\
    $\sigma$& $0.1336 - 0.0205i$ & $0.1336(1) - 0.0205(1)i$  & $0.1334 - 0.0206i$\\
    $\sigma'$& $1.1107 - 0.1701i$ & $1.1105(1) - 0.1697(1)i$& $1.1081 - 0.1702i$ \\
    $\partial\sigma$ & $1.1336 - 0.0205i$ & $1.1334(1) - 0.0203(1)i$ & $1.1328 - 0.0206i$  \\
    $\varepsilon$ & $0.4656 - 0.2245i$ & $0.4656(1) - 0.2240(1)i$  &$0.4633 - 0.2240i$ \\
    $\partial\varepsilon$ & $1.4656  - 0.2245i$ & $1.4650(1) - 0.2207(5)i$ & $1.4623 - 0.2218i$ \\
     $\varepsilon'$  & $1.9083 - 0.5987i$ & $1.9081(2) - 0.5964(2)i$ & $1.8998 - 0.5977i$
    \end{tabular}
    \caption{The central charge and conformal dimensions of the lowest-lying operators of the $Q=5$ Potts CCFT. From left to right: the theoretical predictions \cite{Gorbenko2}, the values from DMRG or QPA fitted to Eq. \eqref{eq:E0Ensubleading} for $\lambda_{c} = 0.0788 + 0.0603i$ for system sizes $L = 8-14$, and the values determined by \cite{Tang2024} at $\lambda_{c}^\text{prev} = 0.079 + 0.060i$ and system sizes $L = 8-13$ using exact diagonalization. } 
    \label{tab:conformaldata_comparison}
\end{ruledtabular}
\end{table} \\ 
\indent \emph{RG flow—}One of the trademarks of complex conformal field theories is their particular renormalization group flow. For the 5-state Potts model specifically, one-loop calculations predicted that the topology of the RG flow resembles a logarithmic spiral \cite{Gorbenko2}. Using the cost function proposed in \cite{Tang2024}, 
\begin{align}\label{eq:costfunction}
    &J(v, g_{\varepsilon'}) = \sum_{n \in \{\phi,\partial \phi \}} \abs{\delta_{\Delta_n} - \delta E_n},  \\
        & \nonumber \delta E_{\phi} = 2\pi g_{\varepsilon'} C_{\phi \phi \varepsilon'}, \quad
    \delta E_{\partial \phi} = \delta E_{\phi} \left(1 + \frac{\Delta_{\varepsilon'}(\Delta_{\varepsilon'} - 2)}{4\Delta_{\phi}}\right),
    \end{align} 
    we systematically recover the parameter $g_{\varepsilon'}$ for different system sizes and $\lambda$ values in theory space \cite{supplementary,reinicke1987}. Here, we also account for the subleading finite-size effects on the energies by using the fits shown in \autoref{fig:lambdacscaling}. The beta function $\beta(g_{\varepsilon'})$ as a function of system size is visible in \autoref{fig:rgflowplot}. The arrows represent the movement of this coupling parameter under RG, which in our case is equivalent to an increase in system size. We recognize the theoretical cyclic behavior of the RG flow. Close to the fixed points, where the conformal perturbation regime holds the best, we see the spiral flow of $g_{\varepsilon'}$. At the fixed points themselves, marked with blue dots, we observe the expected near-zero flow. Unlike in the $O(n>2)$ model \cite{haldar2023}, no separatrix was found between the spiral flow and the high-temperature regime. This is possibly due to the limited grid of the $\lambda$ parameter.\\
\indent We must note that the flow of the values on the real axis, denoted by red arrows, does not reside on the real axis. This is because we plot the conformal perturbation parameter $g_{\varepsilon'}$ instead of the parameter $\lambda_w$ associated with the walking regime. This is also evident from the fact that the unwinding behavior is centered around zero for both $\mathcal{C}$ and $\overline{\mathcal{C}}$ instead of $\lambda_c$ and $\overline{\lambda}_c$. Nonetheless, the RG flow already shows some walking behavior, as seen by the increasing flow farther away from the fixed point. Recovering the walking parameter $\lambda_w$ is a known procedure in the (2+0)D Potts model \cite{Gorbenko2}, but it is not so straightforward in the (1+1)D quantum context. In the supplementary material, we provide further study on this mapping and retrieve the theoretical flow \cite{Gorbenko2}. 

\emph{Complex entanglement spectrum---}To study the complex-valued entanglement spectrum, we make use of the proposed non-Hermitian density matrix \cite{Shimizu2025},
\begin{equation}\label{eq:nonhermrho}
    \rho_{RL} = \frac{\ket{\psi_L} \bra{\psi_R}}{\braket{\psi_R | \psi_L }}, 
\end{equation}
where we identify $\ket{\psi_L} = \ket{\psi_R}^*$ due to the transpose symmetry of the Hamiltonian. In particular, we analyze the entanglement Hamiltonian $K_\mathcal{C}$ \cite{Cardy2016} of the CCFT, which we propose remains related to the density matrix as
\begin{equation}\label{eq:entH}
    \rho_\text{RL} = e^{- 2 \pi K_\mathcal{C}}.
\end{equation}

\begin{figure}[t!]
    {\includegraphics[trim={0cm 0.4cm 0cm 0.5cm},clip,width=0.48\textwidth]{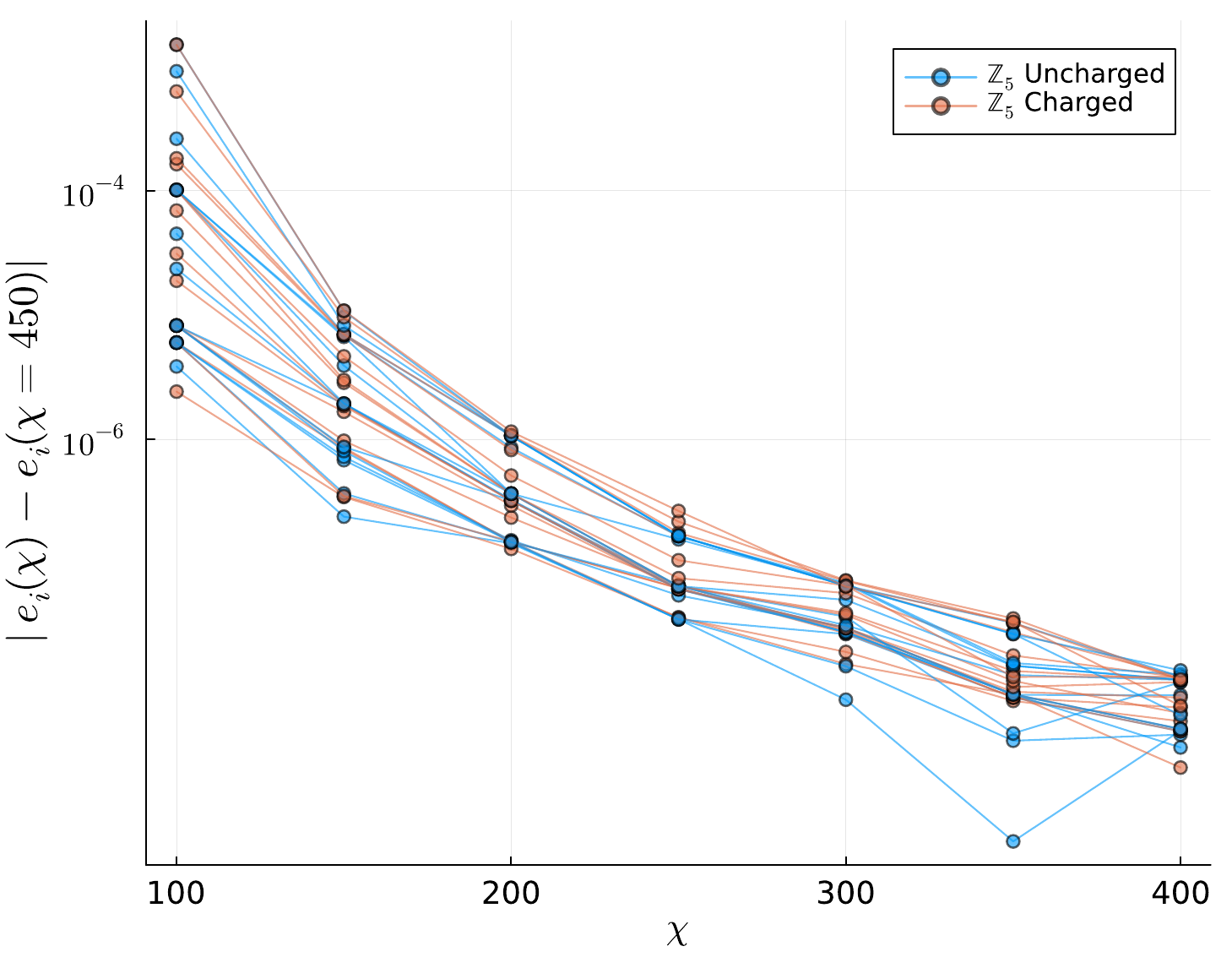}\label{fig: dif ent}}
     \makeatletter\long\def\@ifdim#1#2#3{#2}\makeatother
    \caption{The fifteen largest eigenvalues $e_i$ of the $\mathbb{Z}_5$ charged and non-charged sectors of the reduced density matrix $ \rho^{\mathcal{A}}_{RL}$ for a half-chain bipartition, normalized such that the first eigenvalue of each sector is one, converging for increasing bond dimension $\chi$ to the reference values $e_i(\chi = 450)$. }
    \label{fig: dif ent}
\end{figure} 
 
Through the Bisognano-Wichmann theorem, the Hermitian entanglement Hamiltonian is known to be related to an integral over the Hamiltonian density \cite{Bisognano1975}. Consequently, it has been shown both analytically and numerically that the entanglement spectrum of this entanglement Hamiltonian in the bulk CFT relates to the energy spectrum of the CFT on the boundary \cite{Ohmori2015,Lauchli2013}. The entanglement spectrum itself is obtained from the reduced density matrix $\rho^\mathcal{A}$, which we construct from a non-Hermitian density matrix as \cite{TangMPS2025}
\begin{equation}
    \rho^{\mathcal{A}}_{RL} = M^{\mathcal{B}} C_{R} (M^{\mathcal{A}})^T C_{L},
\end{equation}
with the singular value decompositions $\ket{\psi_R} = \sum_i C^{ii}_{R}\ket{\psi_R^{ i \mathcal{A}}}\ket{\psi_R^{i \mathcal{B}}} $, $\ket{\psi_L} = \sum_i C^{ii}_{L}\ket{\psi_L^{ i \mathcal{A}}}\ket{\psi_L^{i \mathcal{B}}}$ and the rotation matrices $M^{\mathcal{A}}_{ij} = \braket{\psi_L^{i \mathcal{A}} | \psi_R^{j \mathcal{A}}}$ and $M^{\mathcal{B}}_{ij} = \braket{\psi_L^{i \mathcal{B}} | \psi_R^{j \mathcal{B}}}$.  
In Fig.~\ref{fig: dif ent} it is visible how the entanglement spectrum is remaining stable for increased bond dimensions. As further discussed in~\cite{supplementary}, this is due to the fact that for small non-Hermitian deformations, the singular values of the MPS retain the same ordering as the true entanglement structure, explaining why the MPS representations obtained through DMRG and the Quasiparticle Ansatz (QPA) remain robust.

For large enough system sizes, the entanglement spectrum of the lattice with PBC is expected to flow to the free-free conformal boundary condition (CBC), since it is the only one that respects the $S_Q$ symmetry of the Hamiltonian. The boundary CCFT operator content for the free-free CBC has been found in \cite{Tang:2025bju} through finite-size scaling. In particular, three conformal towers with primaries $\mathbb{I}$, $\phi_{3,1}$, and $\phi_{5,1}$ were confirmed with degeneracies one, four, and eleven, respectively \cite{SALEUR1989591}. For our simulations, where we impose the $\mathbb{Z}_5$ symmetry on the MPS, we expect the identity and $\phi_{3,1}$ operator to be located in the zero sector and four charged sectors of $\mathbb{Z}_5$, respectively. The final low-lying operator can be uniquely divided into a three-fold degenerate operator in the zero sector and a two-fold degenerate one in the four charged sectors, as dictated by its possible $S_5$ restriction to $\mathbb{Z}_5$  This eleven-dimensional operator most likely couples to both a five- and the six-dimensional irreducible representation (sector) of $S_5$ as those indeed yield the observed restriction to the sectors of $\mathbb{Z}_5$. \\
\indent \autoref{fig:BCFT_entanglementspectrum} shows the theoretical scaling dimension of these operators and their degeneracies. We find that the logarithm of the eigenvalues of the half-chain density matrix $\rho^{\mathcal{A}}_{RL}$ indeed yields the entanglement spectrum. The entanglement spectrum largely agrees with the free-free boundary CFT spectrum. For example, we identify the degeneracies of $\phi_{5,1}$ with its scaling dimension $\Delta_{\phi_{5,1}} = 4 - 0.9190i$, as there are three indistinguishable levels in the zero sector and two in the one charged sector. Only the imaginary part of the entanglement spectrum seems to converge more slowly. This could be caused not only by MPS limitations \cite{supplementary}, but also because $\rho_{RL}$ is not positive semi-definite, making the relation between the entanglement Hamiltonian and the density matrix less apparent. However, since both the scaling dimensions and the degeneracies are exactly as predicted in \cite{Tang:2025bju}, we conclude that Eq. \eqref{eq:entH} yields information about the boundary CCFT. The entanglement spectrum on OBC will, although less explicitly, also converge to the free-free boundary CFT \cite{supplementary}. \\
\indent\emph{Summary and outlook—}We studied the complex fixed point of the non-Hermitian deformed 5-state Potts model by finite-size scaling and tensor network simulations. We found that the latter can overcome the slightly non-Hermitian nature of the CCFT and converge to trustworthy eigenvalues for system sizes larger than ED. 
We exploited this to refine the estimate of the critical parameter to one additional digit, yielding $\lambda_c = 0.0788 + 0.0603i$. Moreover, using subleading finite-size effects, we obtained more accurate estimates of the conformal data. We also found the renormalization group flow of the coupling parameter $g_{\varepsilon'}$ to exhibit a spiral behavior, in agreement with the theoretical prediction of Gorbenko et al. \cite{Gorbenko2}. Finally, we studied the entanglement spectrum of the ground state at complex criticality. The spectra are found to match those of the boundary CFT with appropriate boundary conditions up to numerical ambiguity in the imaginary part. This phenomenon is well-known for real CFTs due to conformal invariance in spacetime. Our findings provide further evidence of emergent conformal invariance on the lattice for non-Hermitian critical phenomena. \\ \indent
Given the universal scaling behavior of the complex fixed points, a natural question to ask is whether we can similarly establish the finite-entanglement scaling. Simulating non-Hermitian models involves significant numerical challenges, particularly when using periodic boundary conditions for MPS simulations. In this context, the development and application of infinite-size tensor network algorithms, along with their associated scaling techniques, become beneficial. We leave this direction for future work. \\ \indent 
Another interesting question to raise concerns the fascinating physics behind the N\'eel-Valence Bond Solid transition and its corresponding deconfined quantum critical point. As there is no consensus as to whether this transition is weakly first-order or continuous, one can further explore the potential existence of a complex fixed point. We expect that analyzing the entanglement properties through tensor network methods will be powerful for detecting hidden criticality as it was for the 5-state Potts model.\\ \\
\indent \emph{Acknowledgments---}
 The authors thank Wei Tang, Slava Rychkov, Bram Vancraeynest-De Cuiper, and Jutho Haegeman for insightful discussions. B.D. and A.U. were supported by BOF-GOA (Grant No. BOF23/GOA/021). This work was supported by EOS (grant No. 40007526), IBOF (grant No. IBOF23/064), and BOF- GOA (grant No. BOF23/GOA/021). A part of the computation is done on the UGent HPC of the Flemish Supercomputing Centre (VSC).
 \vspace{0.3cm}
 \begin{center}
 \noindent\rule{0.16\textwidth}{1pt} 
\end{center}
\vspace{0.3cm}
\putbib[apssamp]
\end{bibunit}
\clearpage

\onecolumngrid

\begin{center}
\textbf{\large Supplementary Material}
\end{center}

\setcounter{equation}{0}
\setcounter{figure}{0}
\setcounter{section}{5}

\renewcommand{\thesection}{\Alph{section}}
\renewcommand{\theequation}{S.\arabic{equation}}
\renewcommand{\thefigure}{S\arabic{figure}}
\renewcommand{\thetable}{S\arabic{table}}
\renewcommand{\sectionautorefname}{Section \Alph{section}}
\renewcommand{\figureautorefname}{Figure}

\begin{bibunit}[apsrev4-2]
  \tableofcontents
\section{A.$\quad $Extension Term}
\label{app:extensionterm}
In this section, we investigate the form of the extra terms that would allow the RG flow of the 5-state Potts theory to be perturbed onto the CCFT with a complex tuning parameter $\lambda$. Based on the fact that this perturbation will need to be nearly relevant and using properties of the $Q \leq 4$ regime, the perturbation is conjectured to be associated with $\varepsilon'$, the second energy operator \cite{Gorbenko2}. However, since the explicit lattice representation of this operator is not known, a different term is used. The CCFT must respect the same symmetries and dualities as the real fixed point from which it comes, so we expect the extra term to satisfy the $S_Q$ symmetry, translation symmetry on periodic boundary conditions and Kramers-Wannier (KW) duality of the Potts model. Such a term was proposed by Tang et al. \cite{Tang2024}
\begin{align} \label{eq:extension}
    \hat{H}_1 = &\sum_i \sum_{k_1=1,k_2=1}^{Q-1} [ (\hat{X}_i^{k_1} + \hat{X}_{i+1}^{k_1})(\hat{Z}_i^{\dag} \hat{Z}_{i+1})^{k_2} + (\hat{Z}_i^{\dag} \hat{Z}_{i+1})^{k_1} (\hat{X}_i^{k_2} + \hat{X}_{i+1}^{k_2}) ]. 
\end{align}
We illustrate that this term is the unique possible nearest-neighbor extension term that satisfies the above conditions. This is done by considering all possible terms in the local interaction $\mathcal{H}_{i,i+1}$. Using invariance under the KW duality action \cite{Cobanera2010}, one restricts the possible operators to $\{ \hat{X}_i^{k_1}, (\hat{Z_i}^{\dag} \hat{Z}_{i+1})^{k_1},\hat{X}_i^{k_1} (\hat{Z}_i^{\dag} \hat{Z}_{i+1})^{k_2}, \hat{X}_{i+1}^{k_1} (\hat{Z}_i^{\dag} \hat{Z}_{i+1})^{k_2}\}$, up to conjugations under the Weyl algebra. The first two terms are present in the usual Potts Hamiltonian, and therefore must not appear. Using $S_Q$ and translation symmetry, we can show that all remaining terms have the same coefficient. \newline

\noindent Many phenomenological models, such as the Potts model, only contain nearest-neighbor interactions since these are most relevant from the RG perspective, and therefore contribute the most to the RG flow. However, when allowing next-to-nearest-neighbor interactions, other extension terms that obey the symmetries are well-founded. For example, another self-dual, $S_Q$- and translation symmetric extension term is found by generalizing a self-dual Ising term \cite{Alcaraz_2016}, 
\begin{align}
    H_2 = \lambda \sum_i \sum_{k=1}^{Q-1} \left[ (\hat{X}_i \hat{X}_{i+1})^k + (\hat{Z}_i^{\dag} \hat{Z}_{i+2})^k \right].
\end{align} 
The $S_Q$ and translation symmetry can be seen in its matrix representation, whereas the KW duality can be proven using its explicit representation $\phi_d$ in \cite{Cobanera2010}. 
Until now, the analysis of this extended model has been limited to the Ising case ($Q=2$) where it served as a proxy for the perturbation due to the finite bond dimension of a tensor network simulation \cite{Huang2024,Schneider2025}.

\section{B.$\quad $Fixed Point Precision}
\label{app:fixedpointprecision}
The original fixed point of the 5-state Potts Hamiltonian deformed with Eq. \eqref{eq:extension} that corresponds to the CCFT was found in \cite{Tang2024} by means of conformal perturbation theory (CPT) and exact diagonalization. Here, we discuss the different methods that can be used and argue that our critical point, found through subleading finite-size scaling $\lambda_c = 0.0788 + 0.0603i$, is the most accurate. First, it is worth noting that the tensor network simulations were saturated for their bond dimensions, thereby eliminating the need to address those perturbations, as discussed above. \newline

\noindent The original method used to determine the fixed point in \cite{Tang2024} utilized the fact that the perturbing operator has been identified as $\varepsilon'$ \cite{Gorbenko2}, allowing the theory close to the complex fixed point described by the CCFT to be given by \begin{align} \label{eq:CPTtotal}
    H&(J=h=1,\lambda) = \:  H_{CFT} - g_{\varepsilon'}\int dz {\varepsilon'}(z)  \\
    &+  g_{T\overline{T}}\int dz O_{T \overline{T}}(z) +  g_{T^2}\int dz \left[ O_{T^2}(z) + O_{\overline{T}^2}(z)\right]. \nonumber 
\end{align} We isolate the energy difference caused by the perturbation parameter $g_{\varepsilon'}$, namely $ \delta E = E_{g_{\varepsilon'}} - E_{g_{\varepsilon'} = ~0}$, where $E_{g_{\varepsilon'}}$ is the energy in Eq. \eqref{eq:CPTtotal} for some particular parameter value ${g_{\varepsilon'}}$.  This energy difference for primaries $\phi$ and their first descendants $\partial \phi$ is known to be \cite{reinicke1987, Lao2023}
\begin{align} \label{eq:CPTEcorrections}
    \delta E_{\phi} = 2\pi g_{\varepsilon'} C_{\phi \phi \varepsilon'}, \quad
    \delta E_{\partial\phi} = \delta E_{\phi} \left(1 + \frac{\Delta_{\varepsilon'}(\Delta_{\varepsilon'} - 2)}{4\Delta_{\phi}}\right).
\end{align} 
Through radial quantization, one relates the quantum spin chain with periodic boundary conditions to the cylinder with radius $L$, and at criticality $g_{\varepsilon'} = 0$, one can use finite-size scaling (FSS) theory. Consequently, we use these FSS expressions up to leading order
\begin{align}\label{eq:E0Ensubleading}
    \begin{aligned}
         \frac{E_0}{L} &= f - \frac{\pi v c }{6} x^2, \\ 
         (E_{n} - E_0) \times L &= 2\pi v \Delta_n,
    \end{aligned}
\end{align}
where $x = 1/L$, $v$ is the speed of light, $f$ is the ground state energy density, along with the perturbation corrections, Eq. \eqref{eq:CPTEcorrections}, to capture the energy behavior around the critical point. This allows us to use the theoretical values for the scaling dimensions $\Delta_\phi$ and OPE coefficients $C_{\phi \phi \varepsilon'}$ \cite{Gorbenko2} to construct a cost function
\begin{align}\label{eq:costfunction}
    J(v, g_{\varepsilon'}) &= \sum_n \abs{\delta_{\Delta_n} - \delta E_n}, \\
    \delta_{\Delta_n} &= \frac{L}{2 \pi v}(E_{n}-E_{0}) - \Delta_n^\text{th},
\end{align} 
where the parameter $g_{\varepsilon'} $ has a minimum in absolute value, namely zero, at the critical value $\lambda_c$. We emphasize that this cost function also determines $v$ and does not rely on the value we determined by extrapolation in the main text. The residuals of the cost function itself are used as a goodness of fit measure; if its residue is uniform throughout the theory space and significantly smaller than the fit parameter $g_{\varepsilon'}$, the search can be considered trustworthy. \newline

\begin{figure*}[tb]
    \centering
    \includegraphics[trim={0 0.5cm 0 0},clip,scale = 0.26]{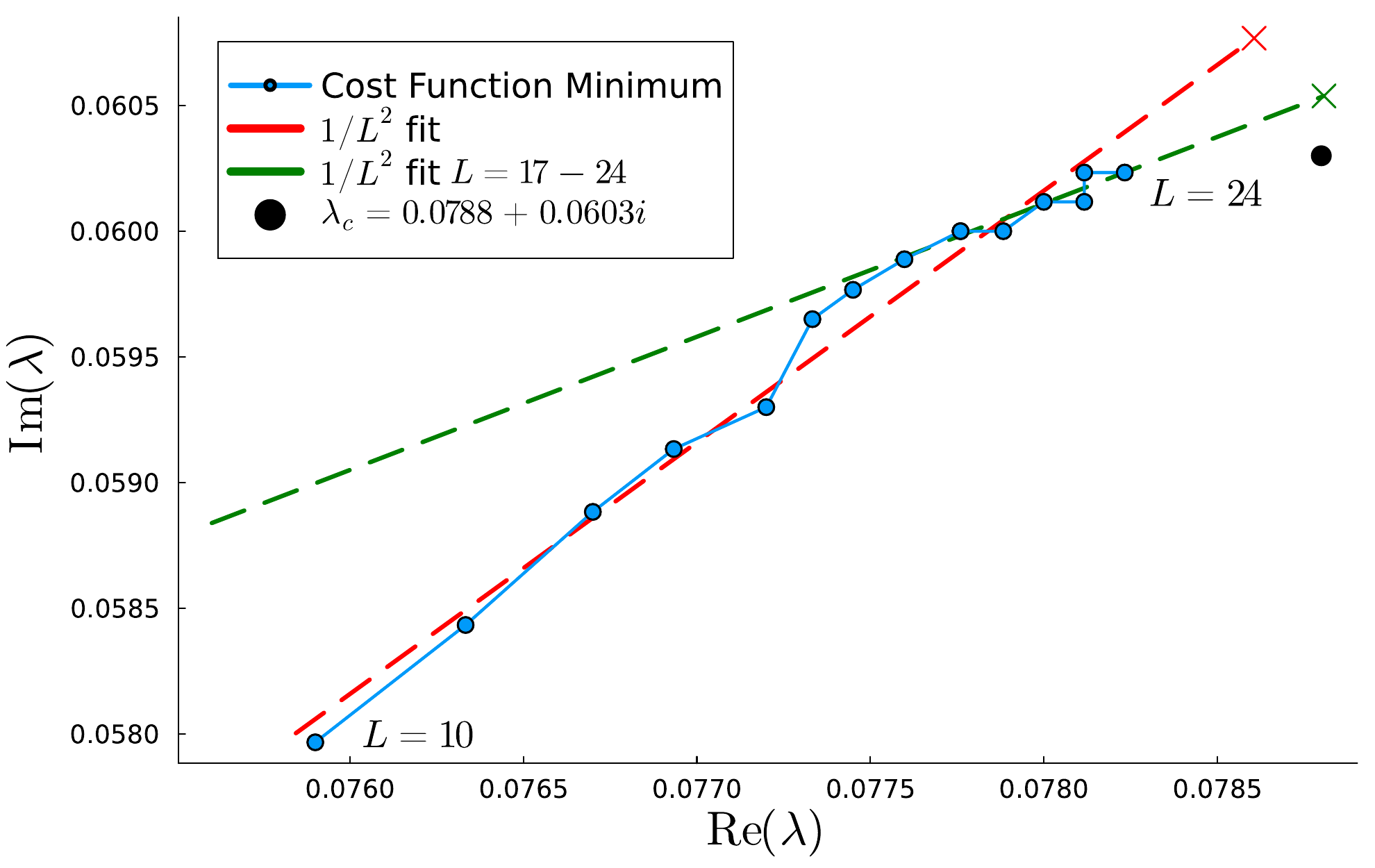}
     \makeatletter\long\def\@ifdim#1#2#3{#2}\makeatother
    \caption{The drift of the critical value due to subleading finite-size corrections on the $g_{\varepsilon'}$ minimum of the cost function, Eq. \eqref{eq:costfunction}, for system sizes $L =10 -24$. We note that the minimum at $L = 18$ and $L = 19$ overlap. We perform a fit of $y = ax^2 +b$ with $x = 1/L$ for all data points and for $L = 17-24$. Their zero points $b$ are marked by a cross and yield an estimate for the critical point, denoted $\lambda_{CF}$. The estimate $\lambda_c = 0.0788 + 0.0603i$ is also included.}
    \label{fig:lambdashift}
\end{figure*}

\noindent The conformal data for the primaries $\sigma$ and $\varepsilon$, together with their first descendants, were used in the cost function, yielding $\lambda_{c}^\text{prev} = 0.079 +0.060i$ \cite{Tang2024}. However, when zooming in around this fixed point, the subleading finite-size effects (FSE) cause the fixed point to move slightly in theory space. This is visible in \autoref{fig:lambdashift}, where we use a more precise grid of $\lambda$ intervals to determine the critical value $\lambda_c$ for system sizes up to $L = 24$. The subleading FSE changes the conformal dimension of the operators $\varepsilon, \sigma,\partial\varepsilon$ and $\partial\sigma$ by a factor which scales as $\sim 1/L^2$. As a result, the cost function will be slightly inaccurate. We observe that subleading FSEs are the source of this inaccuracy, as both the direction of movement and the magnitude of the movement are consistent with the scaling: larger system sizes shift the critical fixed point in a consistent direction, and the displacements decrease with increasing system size. \newline

\begin{figure*}
    \centering
    \includegraphics[trim={0cm 0.4cm 0.2cm 0.4cm},clip,scale = 0.38]{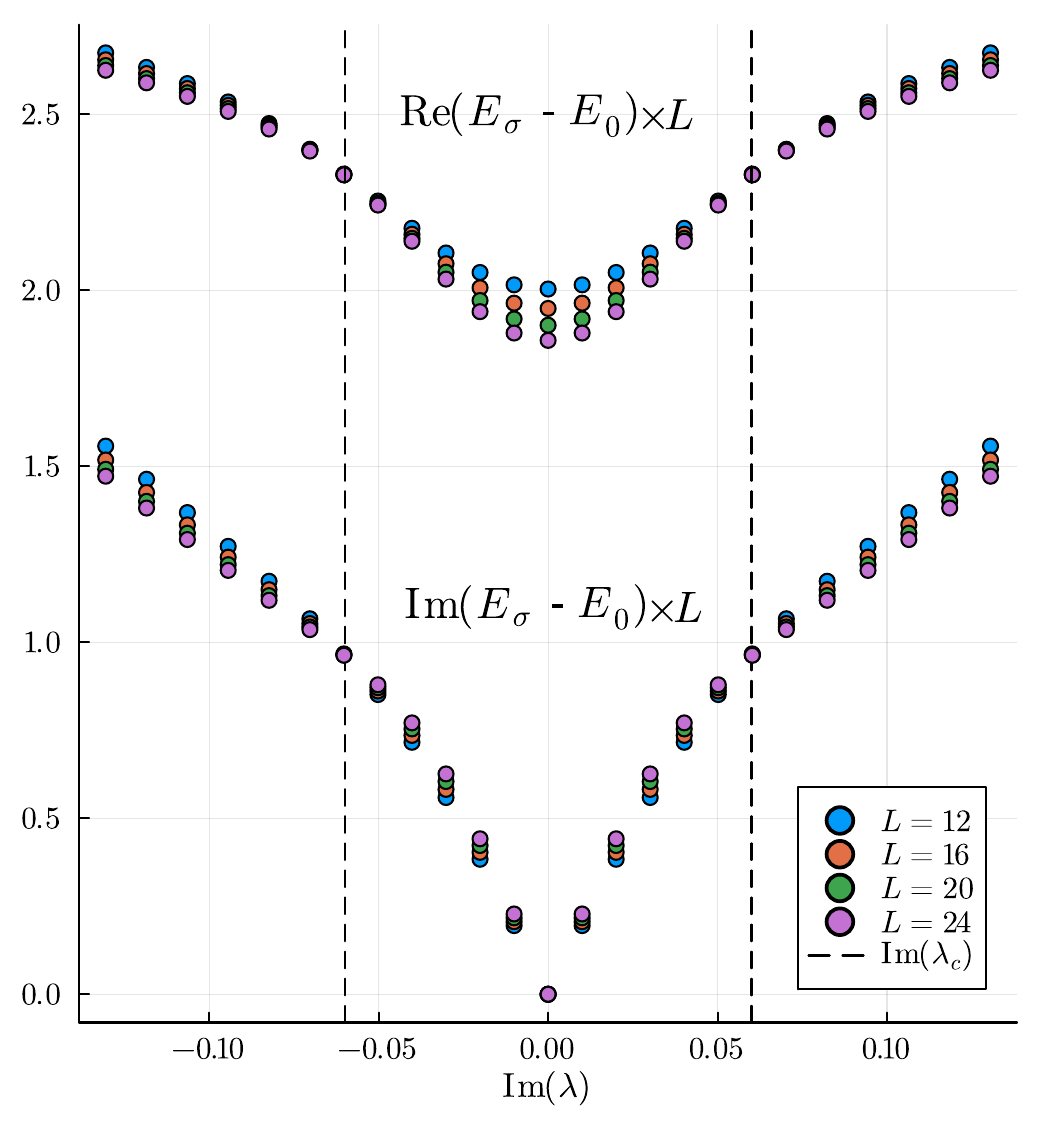}
     \makeatletter\long\def\@ifdim#1#2#3{#2}\makeatother
    \caption{The energy gaps $E_{\sigma} - E_0$ collapsing to $\frac{v \Delta_{\sigma}}{2\pi}$ at the fixed points $\lambda_c = 0.0788 \pm 0.0603i$ when multiplying with different system sizes.}
    \label{fig:datacollapse}
\end{figure*}

\noindent To combat the subleading FSE, we extrapolate the drift to $L \to +\infty$ and determine a more accurate critical point. \autoref{fig:lambdashift} shows this process. By using our data across system sizes $L=10-24$, we find a value of $\lambda_{CF} = 0.0786 + 0.0608i$. However, we find that irrelevant terms with larger scaling dimensions persist for smaller system sizes up to $L\sim 13$. In larger systems, these terms disappear, and the fitting thereof $L=17-24$ yields a different estimate $\lambda_{CF} = 0.0788 + 0.0605i$. 
\begin{figure}[htb]
    \centering
    {\includegraphics[trim={0 0.6cm 0 0},clip,width=0.338\textwidth]{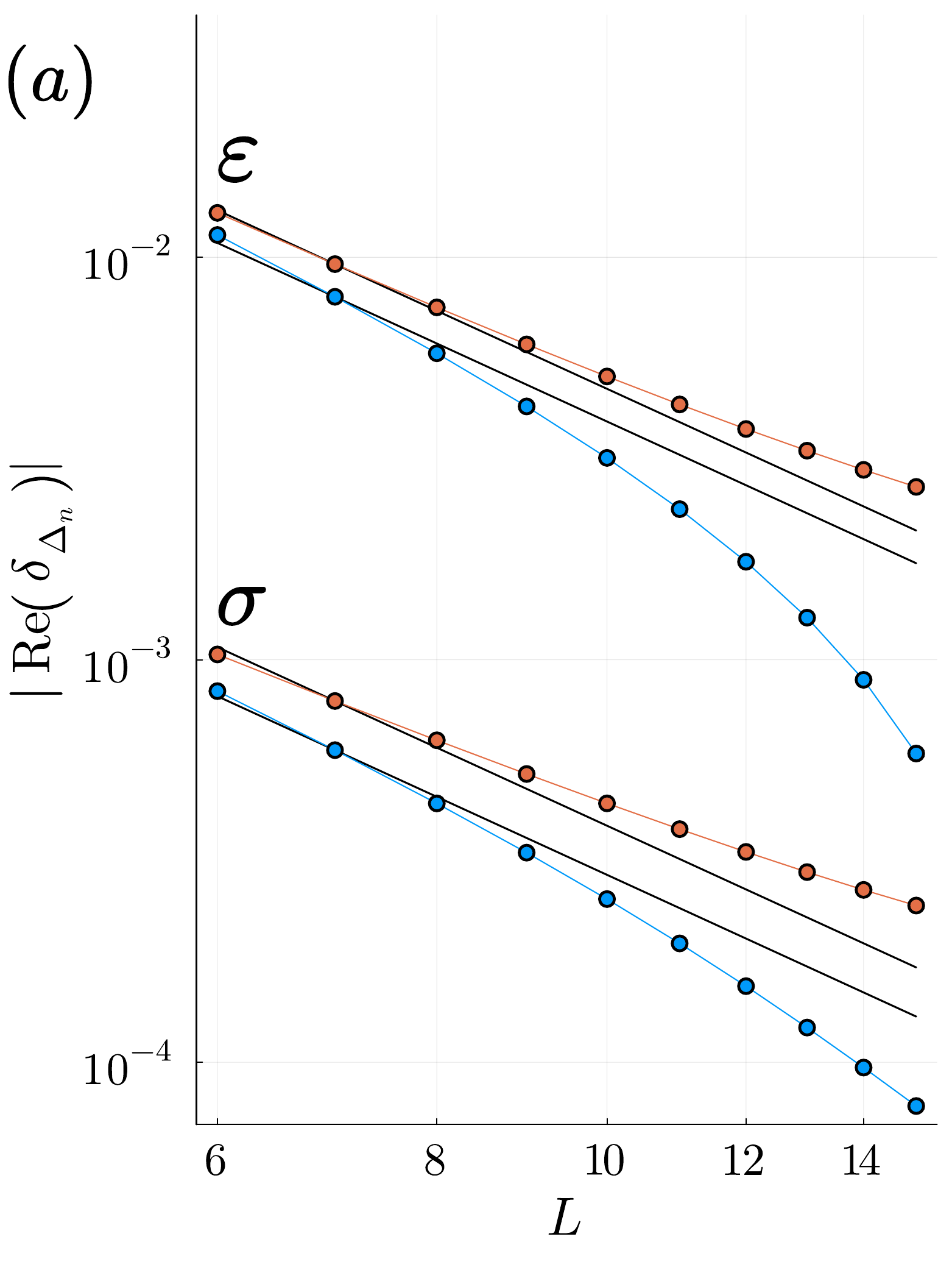}\label{fig:a}}
    \hspace{2cm}
   {\includegraphics[trim={0 0.6cm 0 0},clip,width=0.338\textwidth]{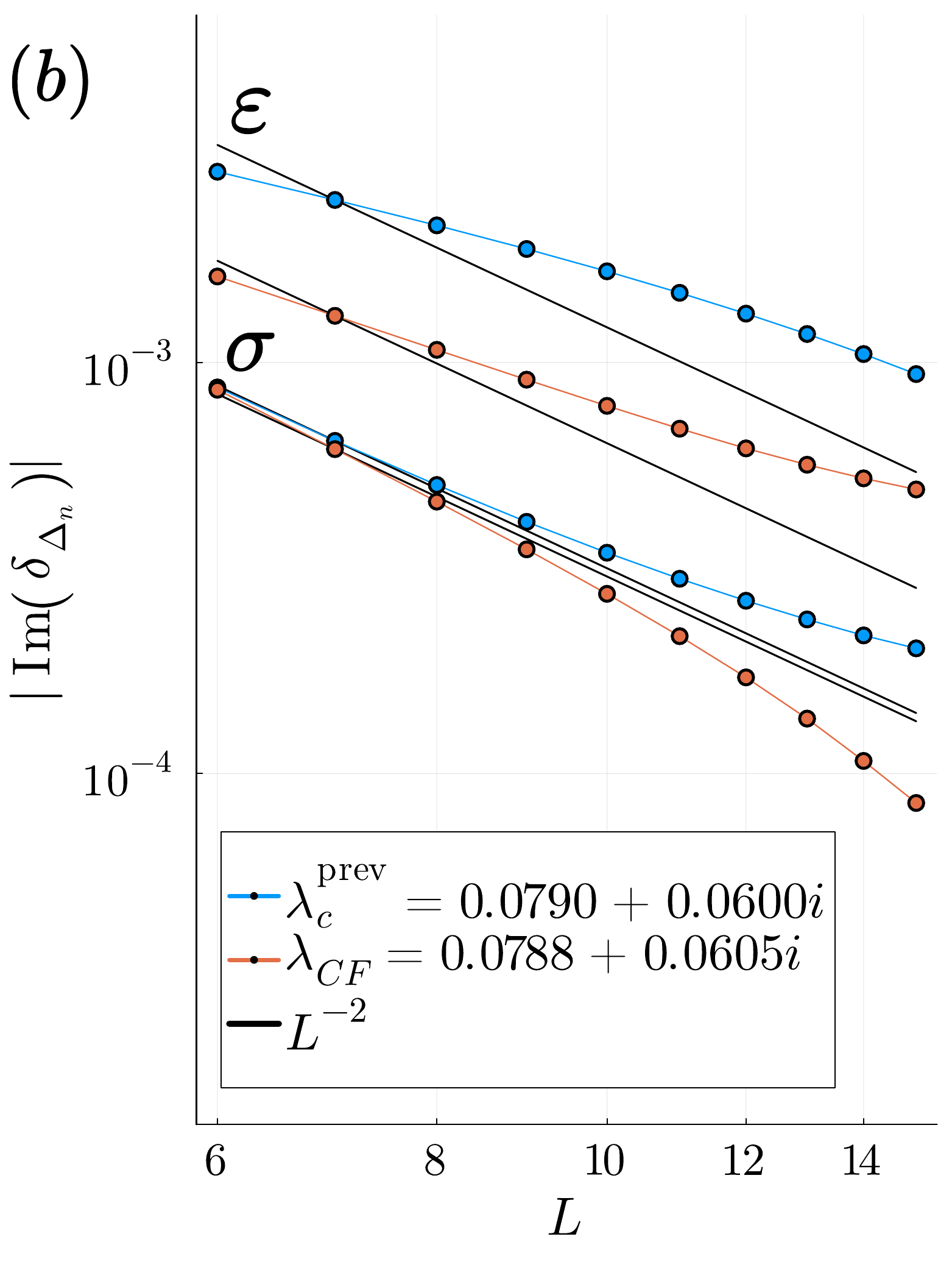}\label{fig:b}}
    \makeatletter\long\def\@ifdim#1#2#3{#2}\makeatother
    \caption{The real (a) and imaginary (b) scaling of the difference between the theoretical and estimated scaling dimensions of $\sigma$ and $\varepsilon$ away from the critical point. $\delta_{\Delta_n}$ is defined in Eq. \eqref{eq:costfunction}.}
    \label{fig:scaling_manylambdas}
\end{figure}
We do not overfit solely by considering the largest system sizes, because in this case, the drift exactly matches the size of the discrete grid used in theoretical space. In \autoref{fig:lambdashift} we also included our newly proposed critical point $\lambda_{c}$, which was determined purely based on finite-size scaling theory with subleading corrections. We see that this is very close to our last estimate of the cost function, with even $\mathrm{Re}(\lambda_{c}) = Re(\lambda_{CF})$. This suggests that the system sizes under consideration are large enough to exclude subleading FSE. The mismatch in the imaginary part remains unclear, but could be attributed to either sub-subleading FSE or the MPS ansatz making less accurate approximations in the imaginary axis, as shown in \autoref{app:sourcesoferror}. \newline

\noindent The downside to either the cost function Eq. \eqref{eq:costfunction} or subleading FSS methods of the main text is that they depend on theoretical predictions of the CCFT data. We demonstrate here that we acquire the same results through data collapse. We show such a data collapse for the $\sigma$ operator in \autoref{fig:datacollapse} over a range of $\mathrm{Im}(\lambda)$ large enough to capture both the CCFT fixed point and its complex conjugate fixed point for $\mathrm{Re}(\lambda) = 0.0788$. We see that both the real and imaginary parts of the energy gaps overlap at the conjugate pair of fixed points, whereas for $\mathrm{Im}(\lambda) = 0$ there is obviously a collapse for the imaginary energy gap; the deformed Potts model recovers its hermiticity. With this, we confirm $\lambda_{c}$ to be the critical value without the need for conformal data of the CCFT. \newline

\noindent In conclusion, we identified three distinct methods that enable us to determine the critical fixed point. We argue that the subleading FSS method from the main text is the most trustworthy, as it incorporates the subleading finite size effects in the most obvious way. Although the other fixed point candidates $\lambda_{CF} =  0.0788 + 0.0605i$ and $\lambda^{\text{prev}}_c = 0.079 + 0.60i$ from \cite{Tang2024} are close to $\lambda_c = 0.0788 + 0.0603i$, the scaling of these other fixed point candidates, acquired by the same method as for $\lambda_{c}$ in the main text, is not consistent. This is shown in \autoref{fig:scaling_manylambdas} where it is visible that the points do not have the correct subleading FSS for the $\sigma$ and $\varepsilon$ operators. This is even disregarding the fact that the system sizes we simulate here only go up to $L$ = 15, contrary to the largest system size $L$ = 24 of the main text. The best scaling is for $\lambda_{CF} = 0.0788 + 0.0605i$, the point that is, in fact, the closest to $\lambda_{c}$. Nevertheless, even though these fixed points are relatively close to $\lambda_{c}$, their subleading scalings are still different. This shows the sensitivity of these subleading FSS to the fixed point accuracy.

\section{C.$\quad $Walking RG Flow}
\label{app:walkingrgflow}
In the main text, we extracted the flow of the coupling parameter $g_{\varepsilon'}$ associated with the CCFT perturbation Eq. \eqref{eq:CPTtotal}. However, this is not the same parameter as the walking parameter $\lambda_w$ associated with the 5-state Potts model \cite{Kaplan2009,Gorbenko2}. Since we suspect that the non-universal value $v$ has to be accounted for and that the transformation will be different depending on which non-Hermitian deformation one is using, it is unclear how to relate the perturbation parameter $\lambda$ in front of Eq. \eqref{eq:extension} back to $\lambda_w$. Moreover, the derivation in \cite{Gorbenko2} was only up to first-order in $\sqrt{Q-4}/{\pi}$. Nonetheless, we expect the transformation between the perturbation parameter $g_{\varepsilon'}$ and the walking parameter $\lambda_w$ to be similar as in \cite{Gorbenko2}, $\lambda_w = x + y \cdot g_{\varepsilon'}$, with complex parameters $x$ and $y$. We find these parameters by demanding that real $\lambda$ values of the extension term yield real values for $\lambda_w$. There is one final degree of freedom, the real part of $x$, related to the start of the RG flow. We fix this by imposing that the critical flows converge to their respective fixed points. The conjugate theory $\overline{\mathcal{C}}$ requires the conjugated parameters $\overline{x}$ and $\overline{y}$. \newline

\begin{figure}[t!]
    \centering
    \includegraphics[trim={
    0cm 0cm 1.5cm 0cm},clip,scale = 0.38]{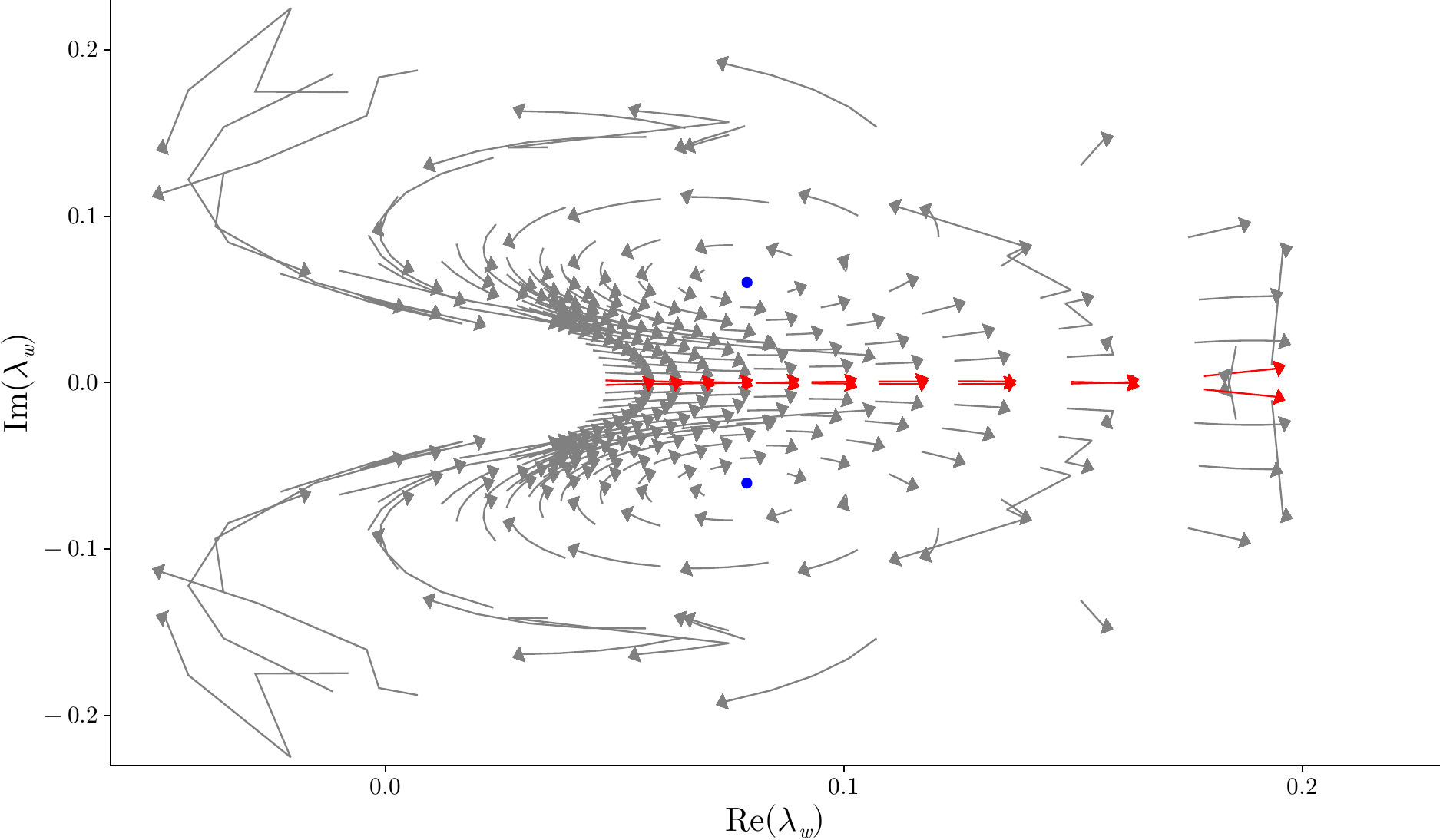}
     \makeatletter\long\def\@ifdim#1#2#3{#2}\makeatother
     \caption{The evolution of $\lambda_w = x +y \cdot g_{\varepsilon'}$ for $L=16-24$  where $x$ and $y$ are fixed such that real theories reside and remain on the real axis. The points corresponding to $\lambda_c = 0.0788 \pm 0.0603$ and the evolution of values on the real axis are in blue and red respectively.}
    \label{fig:RGflow_lambda}
\end{figure}

\noindent We present the resulting RG flow in \autoref{fig:RGflow_lambda}. This figure illustrates, to some extent, the expected topology of the RG flow as presented in \cite{Gorbenko2}. The values that fitted the real parameters on the real axis were $x = 0.076 - 0.015i$ and $y = 0.245-0.057i$. This is, to our knowledge, the first RG flow construction of the deformed 5-state Potts model in a (1+1)-dimensional quantum system. Another CCFT RG flow has previously been identified \cite{haldar2023}. However, this RG flow has a different topology as it is based on a (2+0)D statistical analysis of the O$(n>2)$ model, which is in turn related to the $Q>4$ Potts model. This relation causes its shape to change, not making it comparable to the theoretical predictions of \cite{Gorbenko2}. Regardless, there are areas where the flow seems to deviate in \autoref{fig:RGflow_lambda}. For example, at large $\lambda$ values, it is not even possible to keep real parameters on the real axis. This is most likely due to the failure of the simulation method, as discussed in \autoref{app:sourcesoferror}. Additionally, due to the limited search grid of $\lambda$ used, we did not capture the entire RG flow.

\setcounter{section}{2}
\section{D.$\quad $Open Boundary Conditions}
\label{app:OBC}
\begin{figure}[t!]
    \centering
    \includegraphics[trim={9.8cm 2.5cm 9cm 0cm},clip,width=0.495\textwidth]{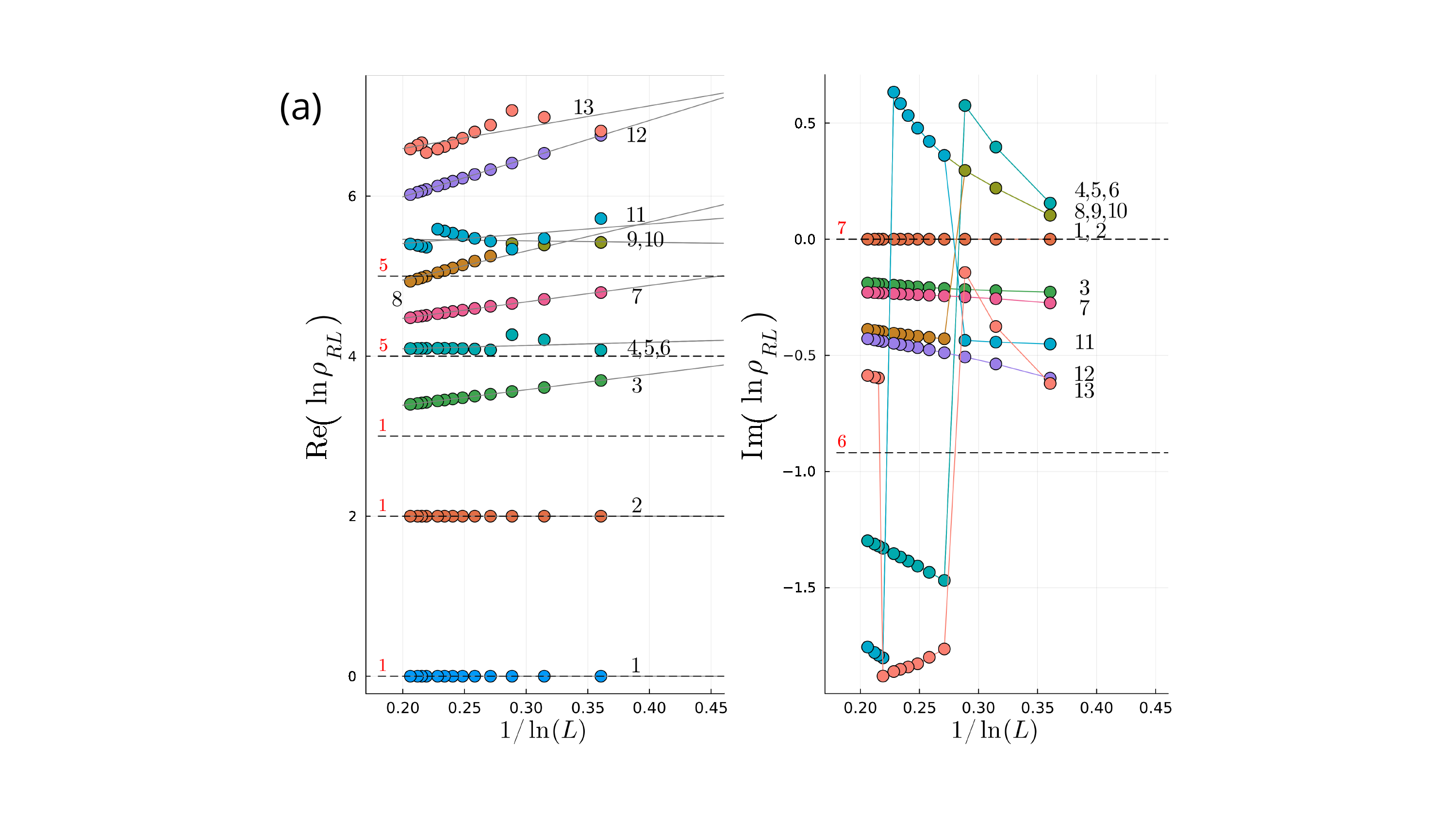}
     \includegraphics[trim={9.8cm 2.5cm 9cm 0cm},clip,width=0.495\textwidth]{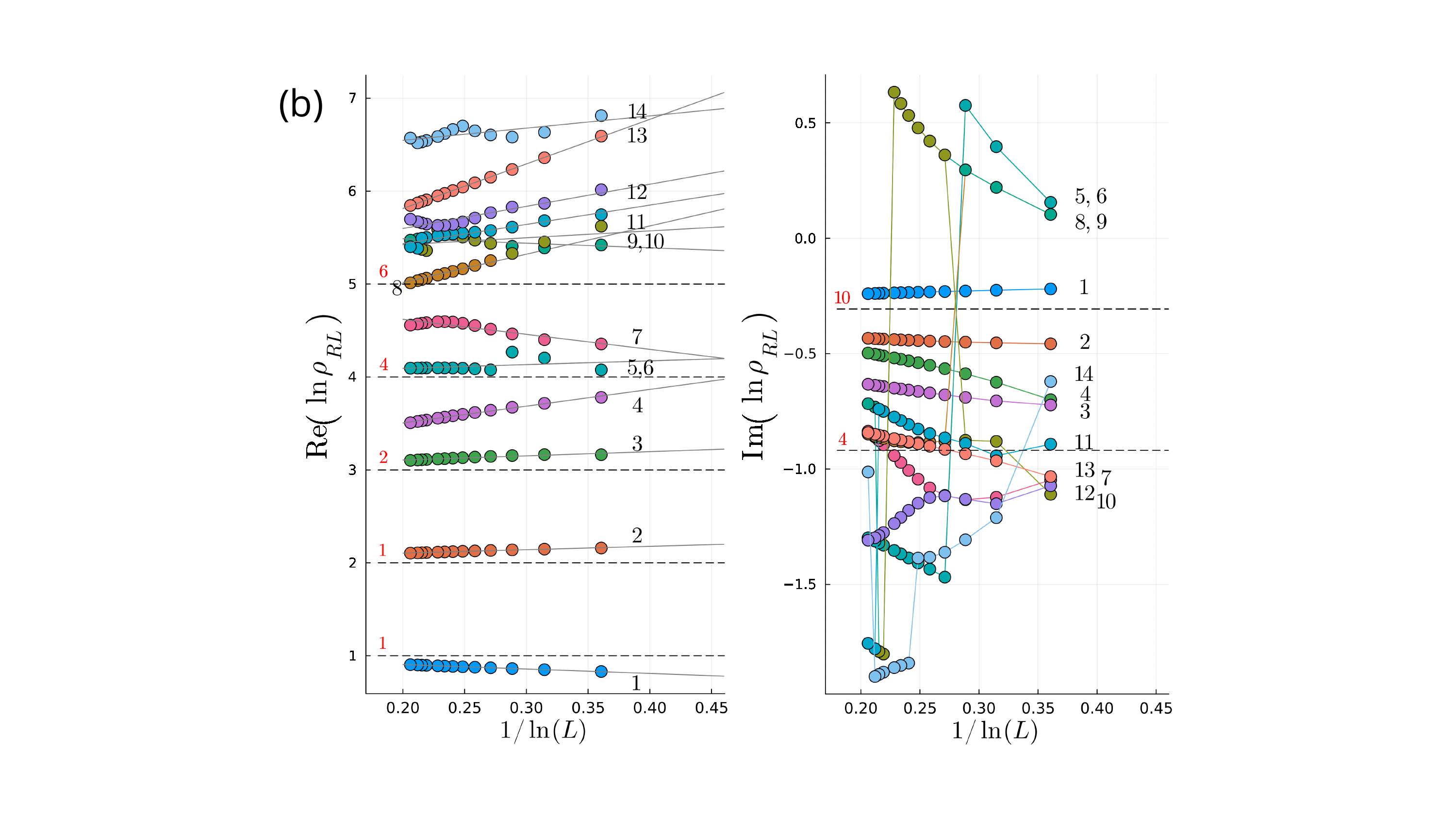}
      \makeatletter\long\def\@ifdim#1#2#3{#2}\makeatother
    \caption{The entanglement spectrum of the ground state in the (a) zero sector and (b) a charged sector of $\mathbb{Z}_5$ for the deformed 5-state Potts Hamiltonian at $\lambda_c = 0.0788 + 0.0603i$ for system sizes $L = 16 - 128$ and open boundary conditions. Here we use DMRG with $\chi = 600$. The dashed lines and red numbers are the theoretical energy levels and their degeneracies, respectively \cite{Tang:2025bju}. We fit only the first two energy levels of the zero sector to their theoretical values.}
    \label{fig:Ent_OBC}
\end{figure}

The entanglement spectrum we determined through the entanglement Hamiltonian on open boundary conditions (OBC) is visible in \autoref{fig:Ent_OBC}. Although it largely agrees with the PBC spectrum presented in the main text, inconsistencies also occur: the real part of the spectrum sometimes diverges from the predicted values, and the imaginary part exhibits what appears to be level crossings for $L \geq 32$. Moreover, we argue that the latter undergoes a potential branch cut in the logarithm, as it drastically jumps from positive to negative values without passing through any level crossings. We attribute these inconsistencies to the unreliability of ground state simulations in capturing the entanglement structure for OBC, as was also observed in the total entanglement scaling \cite{Shimizu2025}. The OBC causes an abrupt end to the lattice, where entanglement cannot flow through, resulting in significant boundary effects. Lastly, the boundary spectrum was, for both PBC and OBC, nearly identical with the previous fixed point estimation $\lambda_{c}^{\text{prev}} = 0.079 +0.060i$.

\section{E.$\quad $Sources of error} 
\label{app:sourcesoferror}
\begin{figure}[htb]
    \centering
    {\includegraphics[trim={0 0.6cm 0 0},clip,width=0.36\textwidth]{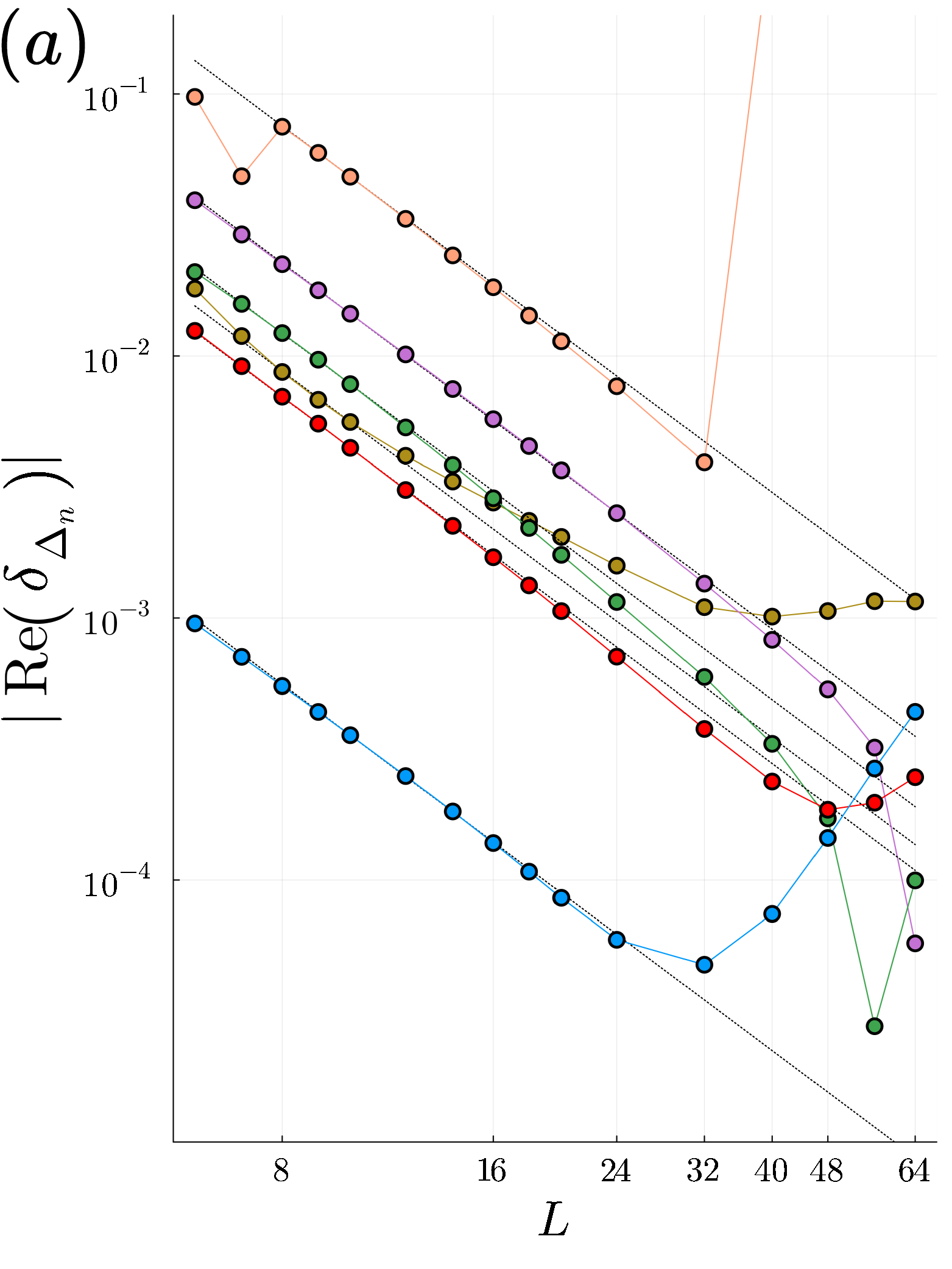}\label{fig:a}}
    \hspace{2cm}
   {\includegraphics[trim={0 2.6cm 0cm 2.4cm},clip,width=0.405\textwidth]{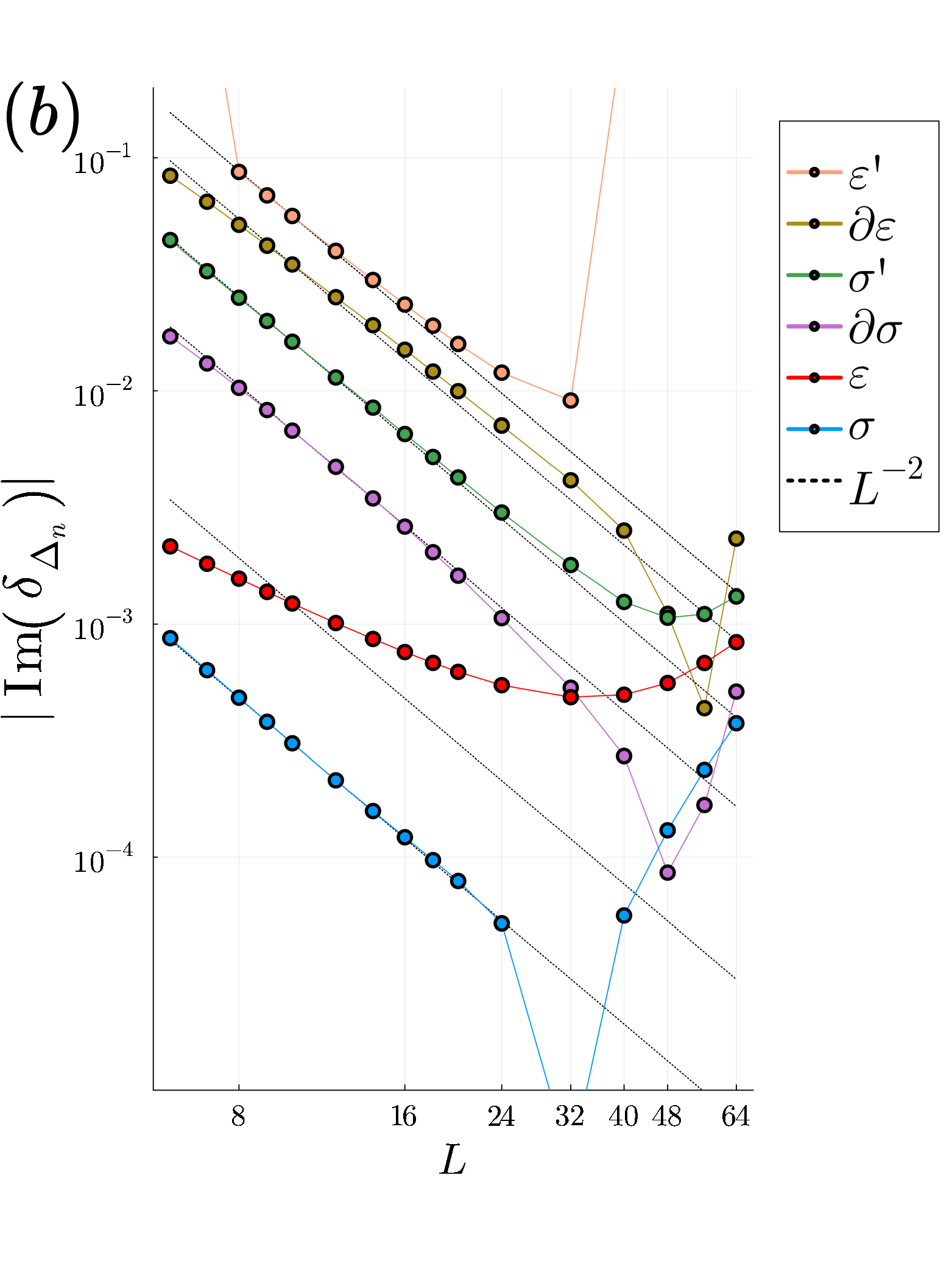}\label{fig:b}}
    \makeatletter\long\def\@ifdim#1#2#3{#2}\makeatother
    \caption{The real (a) and imaginary (b) scaling of the difference between the theoretical and estimated scaling dimensions Eq. \eqref{eq:costfunction} at $\lambda_{c}  = 0.0788 + 0.0603i$ for system sizes $L = 6-64$. The scaling generally agrees with the $1/L^2$ prediction up to large system sizes.}
    \label{fig:correct_fze_longer}
\end{figure}

A large source of errors still arises from the fixed point accuracy. As discussed in \autoref{app:fixedpointprecision}, even though the precision of the fixed point was increased by an additional digit with respect to previous research \cite{Tang2024}, the repulsive nature of the point and its spiral RG flow punishes even small deviations from the true fixed point. Consequently, even for small deviations from the true fixed point, the energy levels will deviate from the expected scaling for large system sizes. We illustrate this in \autoref{fig:correct_fze_longer}, where the deviations from the theoretical scaling dimensions at  $\lambda_{c}  = 0.0788 + 0.0603i$ are plotted for system sizes up to $L = 64$. This repulsiveness is why the system sizes utilized in the predictions of the scaling dimensions are relatively small compared to what is computationally possible. This error is inherent to the nature of complex conformal field theories and can be resolved by repeating the procedure in \autoref{app:fixedpointprecision} for larger system sizes. However, as discussed next, simulation errors prevent this from being done reliably.

\noindent Working within the MPS manifold leads to different sources of simulation errors. First, the MPS representation has a finite bond dimension. This introduces deviations from the true wavefunction because only finite entanglement is present. As a result, there is a perturbation from the true critical point, preventing additional precision of the fixed point \cite{Huang2024}. Moreover, for critical systems, where the ground state's correlation length diverges, and for excited states, which may violate the area law of entanglement, this approximation can be particularly problematic. The limited entanglement at the MPS boundary also introduces additional boundary effects. However, we confirmed that DMRG and QPA reach their bond saturation within (near) algorithmic tolerance for $\chi \geq 400$ and $600$, respectively, for system sizes up to $ L = 24$, regardless of the additional difficulties associated with simulating non-Hermitian models. \newline

\begin{figure}[htb]
    {\includegraphics[trim={12cm 6cm 12cm 6cm},clip,width=0.48\textwidth]{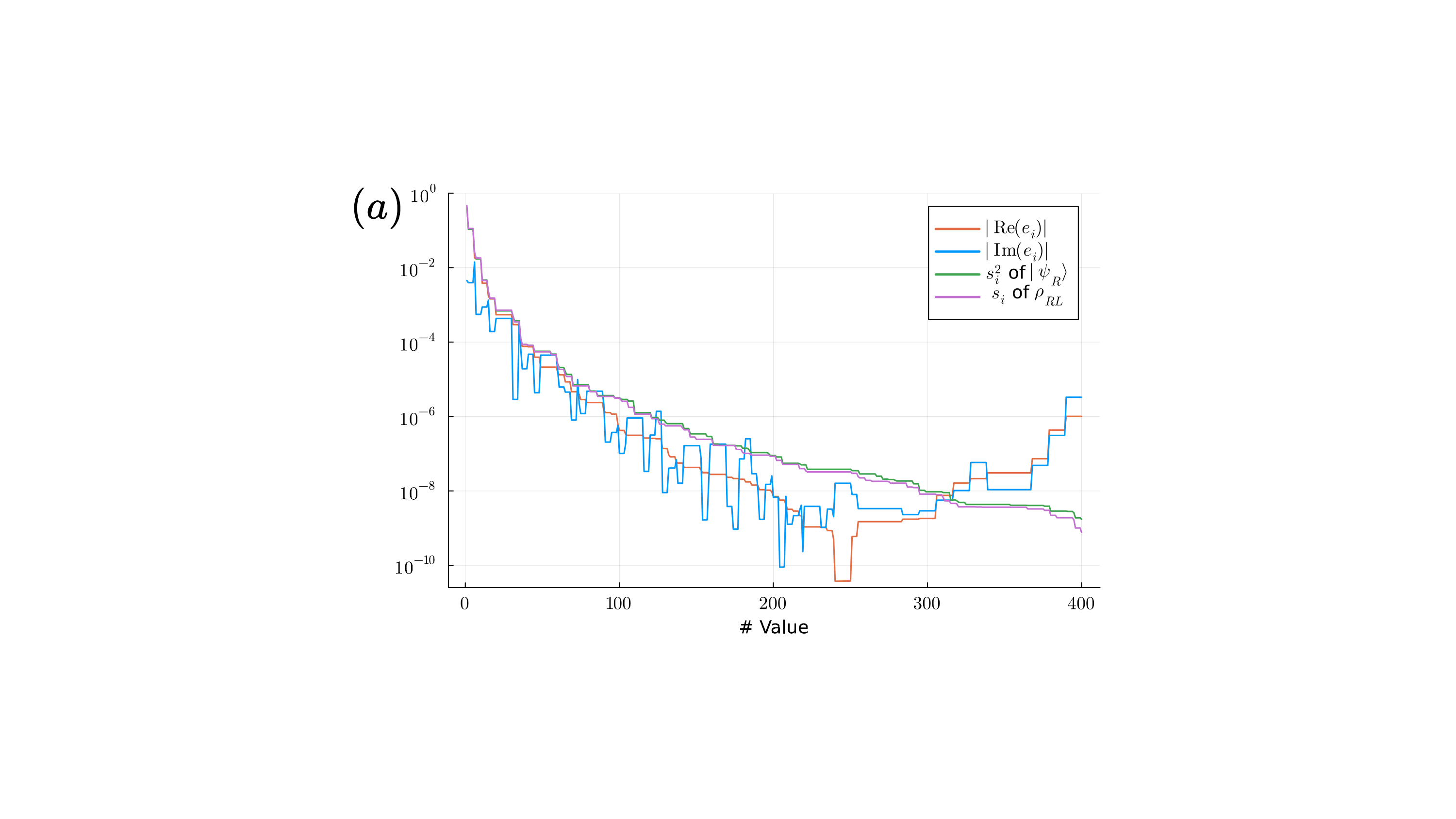}\label{scaling1}}
    {\includegraphics[trim={12cm 6cm 12cm 6cm},clip,width=0.48\textwidth]{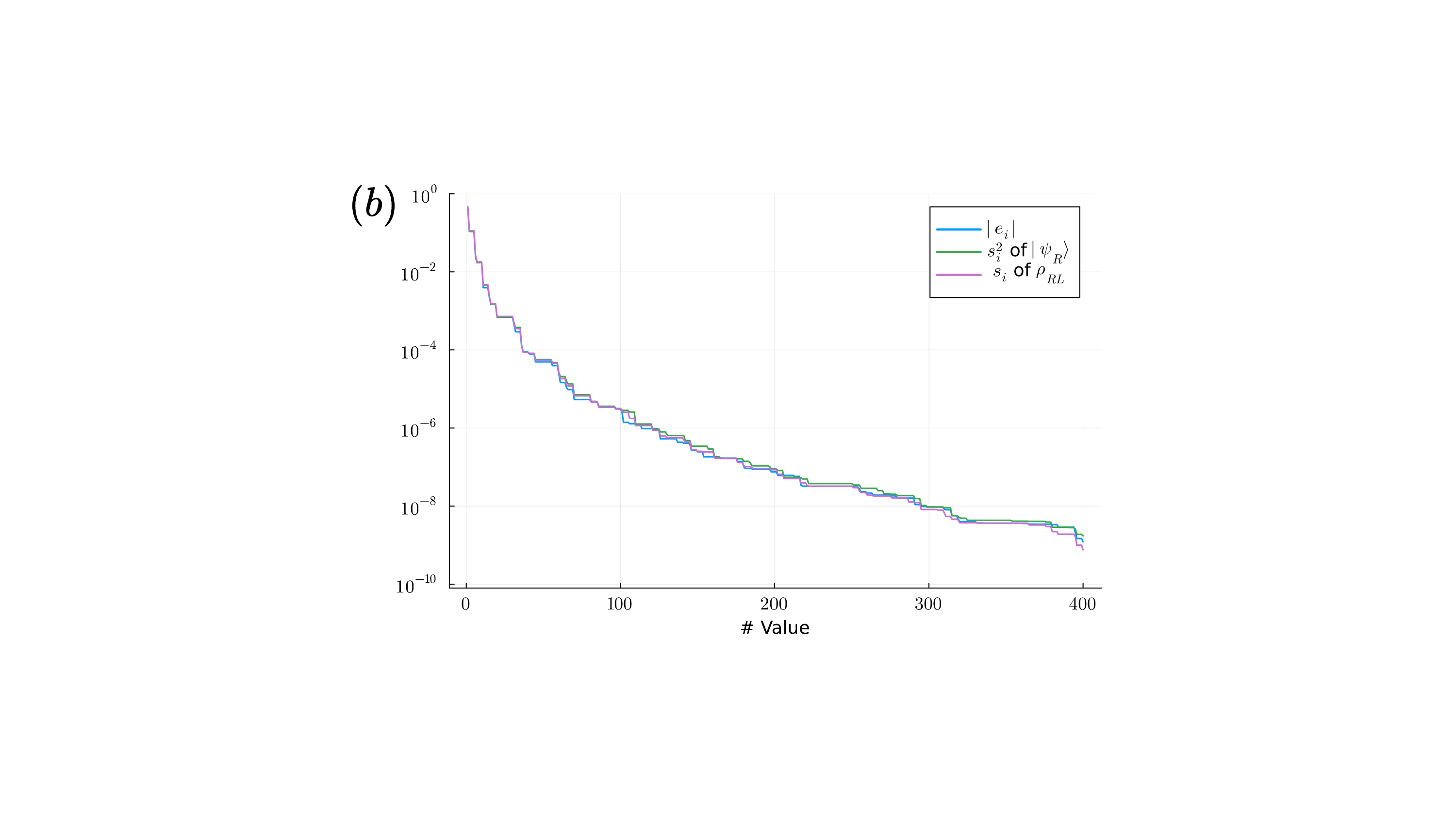}\label{scaling2}}
     \makeatletter\long\def\@ifdim#1#2#3{#2}\makeatother
    \caption{The singular value decomposition $(s_i)$ and (absolute magnitude) of the eigenvalues $(e_i)$ of the non-Hermitian density matrix $\rho_{RL}$ and ground state $\ket{\psi_R}$, ordered by (a) largest real part and (b) largest magnitude, of the deformed 5-state Potts model at $\lambda_c$ in the middle of an $L = 24$ chain.}
    \label{fig:svd_evd}
\end{figure}

\noindent However, even for these smaller system sizes, another limitation may occur. The ordering of the singular values used in the DMRG algorithm is no longer a direct representation of the non-Hermitian entanglement from $\rho_{RL}$. The true entanglement spectrum consists of the eigenvalues $e_i$ of the density matrix, which we calculate for an MPS using the method explained in the main text. To investigate how well this MPS ansatz can find the true entanglement structure of non-Hermitian systems, we look at the difference between the singular values $s_i$ of the ground state wavefunction $\ket{\psi_R}$ and eigenvalues of the corresponding density matrix. In Hermitian systems, these are related by $s_i^2 = e_i$. In non-Hermitian systems, the singular values will remain real, and the eigenvalues can become complex. We show in \autoref{fig:svd_evd} (a) the square of the singular values of the ground state wavefunction $\ket{\psi_R}$, together with the singular and eigenvalues of the density matrix $\rho_{RL}$, of the ground state at the middle of an $L= 24$ chain at the critical point $\lambda_c = 0.0788 + 0.0603i$. These are ordered by largest real part. We also plot the absolute value of the real and imaginary parts of every eigenvalue. Furthermore, this graph is nearly identical throughout the bulk, indicating that the boundary effects have little influence on the entanglement structure. \newline 

\noindent Firstly, it is visible that the squares of the singular values of $\ket{\psi_R}$ and $\rho_{RL}$ are nearly identical. This is because the two are the same up to a normalization constant. More interestingly, we show that the density matrix is not guaranteed to be positive semi-definite as it possesses negative real eigenvalues. We observe this by the increase in absolute value around the value of 250, regardless of its largest real ordering. The largest negative real part of an eigenvalue ends up being of order $10^{-6}$. It is also interesting to note that the ordering according to the largest real eigenvalues corresponds (mostly) to the ordering of the largest imaginary part, indicating some correspondence between the real and imaginary parts. Due to the observed disconnect between singular and eigenvalues, it remains unknown how much entanglement is actually being neglected by the MPS approximation. This is partly rectified by the equivalent ordering property, assuring that if we discard singular values, it will be of similar order in both real and imaginary parts. Since we aim for a certain tolerance in the real spectrum, this requires a certain bond dimension $\chi$. However, as we increase $\chi$, it provides more imaginary contributions to the singular values, which are, relative to their real counterpart, smaller in absolute value. As a result, by imposing a large bond dimension, a larger portion of low, untrustworthy, and imaginary entanglement is retained. This might explain why the imaginary entanglement is known to increase incorrectly with the bond dimension \cite{Shimizu2025}, and why the conformal dimensions and spectra shown in the main text and in \autoref{fig:Ent_OBC} are less accurate in the imaginary part. \newline

\noindent \autoref{fig:svd_evd} (b) further illustrates that when ordering by the magnitude of the eigenvalues, we recover the scaling of the singular values. This might be due to the symmetric nature of the Hamiltonian.
\noindent The figure shows that the MPS ansatz retains singular values corresponding to the absolute magnitude of the entanglement of our system. However, as the imaginary entanglement entropy is smaller relative to its real part but decreases more slowly, a larger number of untrustworthy imaginary values is retained, causing a systematic error. Since higher excited states will have an even flatter entanglement structure, causing more imaginary values close to the cutoff, we expect this approximation to be even worse for excited states. \newline

\noindent For complex conformal theories with larger non-Hermitian contributions compared to their Hermitian counterpart, such as the recently discussed $O(N>2)$ nonlinear sigma models, we do not expect this algorithm to work, as the truncation errors would become too large \cite{yang2026}. In this case the truncation has to be done based on the true entanglement structure of the generalized density matrix. Candidates for such a method include the biorthonormal-block density-matrix renormalization group algorithm or the fidelity algorithm \cite{Zhong2025,Yamamoto(2022)}. \newline

\noindent To recapitulate, in this work we employed the standard DMRG and the Quasiparticle Ansatz to determine the ground state and excited states, respectively \cite{White1992,Haegeman2013}. We found these to agree with ED up to the imposed algorithmic tolerance of $10^{-6}$ for $\lambda_c = 0.0788 + 0.0603i$ and $L = 10$. 
Only when increasing the imaginary part of $\lambda$ in the deformed Potts Hamiltonian do we observe that these algorithms make errors that are multiple orders of magnitude larger than the imposed tolerance. This might be due to the smallest real ordering, used in both algorithms, no longer representing the ground state. Moreover, these algorithms do not account for the differences between left and right eigenvectors. We briefly mention that we also attempted to find excited states via DMRG by projecting out the ground state. We found these to agree with QPA for $\chi \geq 400$, but its convergence was much slower.
\vspace{0.3cm}
 \begin{center}
 \noindent\rule{0.32\textwidth}{1pt} 
\end{center}
\vspace{0.3cm}
\addcontentsline{toc}{section}{References}
\putbib[supp]
\end{bibunit}

\end{document}